\documentclass[11pt,a4paper]{article}
\usepackage[T1]{fontenc}
\usepackage{lmodern,amsmath,amssymb,amsthm,mathtools,microtype}
\usepackage[margin=24mm,headheight=14pt]{geometry}
\usepackage{booktabs,array,enumitem,xcolor,longtable}
\usepackage[colorlinks=true,linkcolor=blue!45!black,citecolor=blue!45!black,urlcolor=blue!45!black]{hyperref}
\newcommand{\dd}{\mathrm d}\newcommand{\atanh}{\operatorname{atanh}}
\newcommand{\AC}{\operatorname{AC}}\newcommand{\Rop}{\mathcal R}\newcommand{\Mop}{\mathcal M}
\newcommand{\B}{\mathcal B}\newcommand{\Id}{\mathsf I}
\newtheorem{lemma}{Lemma}
\allowdisplaybreaks[2]
\usepackage{eso-pic}
\newcommand{\preprint}[1]{%
  \AddToShipoutPictureFG*{%
    \AtPageUpperLeft{%
      \put(\LenToUnit{\dimexpr\paperwidth-1.5cm\relax},
           -\LenToUnit{1.5cm}){%
        \makebox(0,0)[r]{\small #1}%
      }%
    }%
  }%
}
\preprint{QMUL-PH-26-40}
\title{Exact static two-body gravitational potential at first self-force order}
\author{Ricardo Monteiro\\[5pt]
\small Centre for Theoretical Physics, Department of Physics and Astronomy,\\
\small Queen Mary University of London, E1 4NS, United Kingdom}
\date{}
\begin{document}
\maketitle
\begin{abstract}
The static interaction potential of two non-spinning point masses in general relativity was previously computed through seventh post-Newtonian order using methods of effective field theory. Its probe limit, or zeroth-self-force contribution, is determined exactly by the Schwarzschild geometry. We derive a closed-form expression for the static first-self-force correction, quadratic in the light mass, and establish its weak-field expansion to all orders. Working in harmonic gauge, we construct the required Green's function for the coupled linearised gravitational equations about the Schwarzschild background. We fix its homogeneous contributions and regularised coincident limit by matching to the dimensionally continued point-source theory. The result reproduces known perturbative coefficients and determines all higher static post-Newtonian coefficients in this mass sector.
\end{abstract}

\setcounter{tocdepth}{2}
\tableofcontents

\section{Introduction and result}

The two-body problem in general relativity, which plays a key role in gravitational-wave physics, is a formidable mathematical challenge~\cite{Blanchet,LISAReview}. Because the problem has many parameters and spans a wide range of scales, analytical control remains essential wherever it can be achieved, despite the progress of numerical relativity. Three expansions provide such control, each in a different small parameter. The post-Newtonian (PN) expansion assumes weak fields and slow motion~\cite{SchaeferJaranowskiReview}. The post-Minkowskian (PM) expansion is an expansion in Newton's constant $G$ at fixed velocities, so it assumes weak fields but keeps the full velocity dependence. Gravitational self-force theory expands in the mass ratio and keeps the full strong-field dependence. The effective-one-body formalism is designed to combine information from these expansions~\cite{BuonannoDamour,Damour2016}. Both the PN and PM expansions admit an effective field theory (EFT) description of the compact bodies in terms of point particles~\cite{GoldbergerRothstein,Porto,LeviReview,KalinPortoFramework}.

In the PN regime, complementary calculations of the conservative dynamics use harmonic-coordinate equations of motion, Hamiltonian methods and EFT~\cite{Blanchet,SchaeferJaranowskiReview}. Within EFT, systematic power counting organises the calculation into Feynman diagrams, allowing automation and computations at high order~\cite{FoffaSturani3PN,FoffaSturani4PN}. The complete fourth-PN EFT calculation combines potential and radiation contributions, including conservative tail effects, with a consistent treatment of their infrared and ultraviolet singularities~\cite{FoffaPortoRothsteinSturani}.

The PM expansion has been transformed by methods from quantum field theory. See refs.~\cite{BernReview,BjerrumBohrReview,KosowerReview,BuonannoReview} for reviews. These developments include early work on extracting classical potentials from scattering amplitudes~\cite{NeillRothstein,BjerrumBohr2018,CheungRothsteinSolon,Bern3PM,BernDoubleCopy}, the observables-based KMOC formalism~\cite{KMOC}, heavy-mass effective theory~\cite{DamgaardHaddadHelset,HEFT}, worldline EFT~\cite{KalinPortoFramework,KalinLiuPorto}, worldline quantum field theory~\cite{WQFT}, and the gravitational eikonal~\cite{DiVecchiaReview}. Results include conservative dynamics at fourth PM order~\cite{Bern4PM,Dlapa4PMa,Dlapa4PMb}, fifth-PM results at first self-force order~\cite{WQFT5PM,Dlapa5PM}, and the complete fifth-PM potential-graviton contributions for arbitrary mass ratio~\cite{Bern5PM}.

Self-force theory treats the lighter body as a perturbation of the exact black-hole geometry of the heavier one. The field it sources acts back on its motion, through both conservative effects and radiation reaction, which makes the self-force expansion the natural description of extreme-mass-ratio binaries~\cite{PoissonPoundVega,BarackPoundReview,PoundWardell,LISAReview}. EFT and curved-background amplitude formulations now connect it directly to relativistic scattering~\cite{CheungEMRI,CheungEMRIFull,KosmopoulosSolon}, and recent work has developed analytical time-domain scattering waveforms at first self-force order, within a framework that organises their PN and PM expansions to arbitrary order~\cite{ParraSelfForceWaveforms}. A common approach decomposes the field of the small body into multipoles about the black hole and regularises it mode by mode~\cite{BarackOri}. In such calculations, the lowest multipoles, the monopole and the dipole, require separate treatment~\cite{DetweilerPoisson} --- a feature we will also encounter.

The conservative, local-in-time part of the two-body dynamics can be described by a Hamiltonian, whose potential depends, for non-spinning point masses, only on the momenta of the bodies and their separation. In this paper we focus on the static part, which survives when both bodies are at rest. In the EFT, it is the part of the effective action that remains when the velocities are set to zero. This is a well-defined part of the potential, and it does not require the two bodies to be in equilibrium. The static potential had previously been computed through seventh PN order, that is, through $G^8$, with the fourth~\cite{Foffa4PN,DamourJaranowski}, fifth~\cite{Foffa5PN,Blumlein5PN}, sixth~\cite{Brunello6PN} and seventh~\cite{Brunello} orders obtained in the last decade. Most of these calculations use EFT methods, with harmonic gauge and dimensional regularisation for the Feynman integrals, while direct position-space calculations have provided independent checks and corrections~\cite{DamourJaranowski}. Further developments include resummations of high-order PN contributions~\cite{FoffaSturaniTorres}, graph algorithms for worldline EFT~\cite{KalinLim}, and static calculations in scalar--tensor gravity~\cite{AlmeidaLiuZhang}.

With masses $m$ and $M$ for the bodies and harmonic-coordinate separation $R$, the static potential can be expanded as
\begin{equation}
\label{eq:V}
V_{\text{static}} = \frac{GmM}{R}\sum_{a,b\geq0} V_{a,b}\,\left(\frac{G}{R}\right)^{a+b} m^a M^b,
\qquad (c=1)
\end{equation}
where $V_{a,b}=V_{b,a}$ are numerical coefficients, with $V_{0,0}=-1$, such that the post-Newtonian order of each term is $a+b$. Given the relative simplicity of the static potential, one may hope to re-sum certain pieces, which is precisely what we achieve.

In the approximation $m\ll M$, we can write
\begin{equation}
\label{eq:Vsf}
V_{\text{static}} = M\left[\frac{m}{M}\,v_{0SF}(u) + \frac{m^2}{M^2}\, v_{1SF}(u) + O\!\left(\frac{m^3}{M^3}\right)\right], \qquad u=\frac{GM}{R},
\end{equation}
where $v_{nSF}(u)$ encodes the $n$-th self-force order~\cite{PoundWardell}.
The remainder denotes third and higher powers of $m/M$ at fixed $M,R$. 0SF corresponds to the probe limit, governed by the Schwarzschild geometry sourced by $M$. This geometry can also be reconstructed perturbatively from field theory and scattering amplitudes~\cite{Duff,NeillRothstein}, and recent constructions recover it to all orders in $G$~\cite{MougiakakosVanhove,DamgaardLee}. In harmonic coordinates, it is well known to imply
\begin{equation}
\label{eq:v0SF}
v_{0SF}(u) = \sqrt{\frac{1-u}{1+u}}-1.
\end{equation}
\paragraph{Main result.} Writing $\atanh u=\tfrac12\log[(1+u)/(1-u)]$, we obtain
\begin{equation}
\label{eq:v1SF}
v_{1SF}(u) = \frac{u^2\left[1-4u+\frac{4}{3}\,u^2\operatorname{atanh}u\right]}{2(1-u)(1+u)^3}.
\end{equation}
Expanding for small $u$,
\begin{align}\label{eq:v1series}
v_{1SF}(u) ={}& \frac{u^2}{2}-3u^3+6u^4-\frac{31}{3}u^5+\frac{91}{6}u^6-\frac{190}{9}u^7\notag\\
&+\frac{248}{9}u^8-\frac{1574}{45}u^9+\frac{429}{10}u^{10}+O(u^{11}).
\end{align}
The coefficients through $u^8$ match ref.~\cite{Brunello}, and the closed expression determines all higher coefficients in this mass sector. Comparing to \eqref{eq:V}, $V_{1,b}$ is the coefficient of $u^{b+2}$ in $v_{1SF}$.

The resummed expression~\eqref{eq:v1SF} is singular as $u\to1$, which in harmonic gauge corresponds to the horizon of the background Schwarzschild geometry sourced by the heavy mass. The mass-ratio expansion of this static potential therefore breaks down near the horizon: $v_{1SF}$ grows like $(1-u)^{-1}$, up to a logarithm, while $v_{0SF}$ stays finite, so the first-self-force contribution becomes comparable to the probe contribution when $1-u$ is of order $m/M$, up to logarithmic factors. This is a property of the static potential in harmonic gauge, not a general limitation of self-force theory. Our derivation establishes the result coefficient by coefficient in the weak-field expansion, without relying on the near-horizon region. In self-force language, $v_{1SF}$ is the first correction to the probe limit due to the light body's own gravitational field.

We hope that our work will prove useful for future calculations of the full (static and non-static) gravitational potential --- not just due to the closed-form result, but especially due to the method used to derive it.

\paragraph{Outline.}
At first self-force order, the only new ingredient is the linear response to the light body on the exact heavy-body background. In the static case, the light body sources the gravitational scalar (encoding the time--time metric component) but the scalar is coupled to the reduced spatial metric in $d$ spatial dimensions, so both respond. We first decompose these perturbations into angular multipoles. Then, we determine the radial Green's functions with source at the light body, fixing their homogeneous parts as in the standard perturbative calculation with dimensional regularisation (that is, through the point sources at the two bodies, with no separate condition at the horizon). Finally, we take the coincident limit of the radial Green's functions and sum over all multipoles, which again requires regularisation. The paper is organised as follows. 

In Section~\ref{sec:setup}, we recall the set-up of ref.~\cite{Brunello}, which translates the calculation of the static potential into that of connected correlators in a theory with auxiliary sources associated with the two bodies. At first self-force order, the relevant correlators have two insertions of the light source, with the heavy source resummed into the background. Holding the physical heavy mass fixed also shifts the heavy auxiliary source, which contributes to the potential alongside the light-source response.

Sections~\ref{sec:response}--\ref{sec:exceptional} construct the radial Green's functions for each multipole. In Section~\ref{sec:response}, we set up the coupled scalar--metric perturbation problem and fix its homogeneous solutions by matching to the EFT at generic dimension. Section~\ref{sec:continuation} is the heart of the argument: setting $d=3$ too early loses a finite constant, which we retain by continuing one homogeneous solution in the dimension, and this solution then allows us to exclude dimensional poles for $\ell\geq2$. We construct the resulting Green's functions from Legendre solutions in Section~\ref{sec:higher}. As in standard self-force calculations~\cite{BarackOri,DetweilerPoisson}, the monopole ($\ell=0$) and the dipole ($\ell=1$) need separate treatment, which we give in Section~\ref{sec:exceptional}.

Sections~\ref{sec:angular} and~\ref{sec:summation} address the sum over multipoles of the coincident Green's functions. We show that the angular sum may be evaluated at $d=3$ after subtracting a local term whose dimensionally regularised sum vanishes, and then perform the final sum, where most terms telescope or cancel. This yields the correlators, which we then translate into the potential. We conclude in Section~\ref{sec:discussion}. The appendices collect technical results on the radial operators, the dipole and the angular sum, and Appendix~\ref{sec:app_checks} describes the computational checks.

\paragraph{Supplementary material.} The ancillary files contain the code used to check the calculations, together with reference outputs, logs and a guide to reproducing them, starting from \texttt{START\_HERE.txt}. Appendix~\ref{sec:app_checks} explains what these checks cover and how they relate to the proof.

\paragraph{Note added.} While this paper was being completed, Liu and Yuan~\cite{LiuYuan} posted a paper that derives the same closed-form static first-self-force potential, and also determines the full static potential to $G^{12}$. Calculations for their paper and ours were carried out independently. Our Eq.~\eqref{eq:v1SF} agrees with their Eq.~(3), and our dimensionless connected correlators $\overline\Gamma_{2,2n}$, with two light-source and $2n$ heavy-source insertions, agree with their Eq.~(M18). Beyond first self-force order, we also independently computed $\overline\Gamma_{3,7}=2814336$, in agreement with their result.

\section{Auxiliary sources and the static potential}
\label{sec:setup}

Because dimensional regularisation is needed, we must consider the gravitational action in $d+1$ dimensions, where $d$ is the number of spatial dimensions. Its physical value is $d=3$, but the limit $d\to3$ is subtle due to $0/0$ ambiguities, so the limit can only be taken once the ambiguities are resolved. Since we are interested in the static problem, we can start with the symmetry-reduced metric~\cite{KolSmolkin,KolSmolkinGauge}:
\begin{equation}
 ds^2=e^{2\phi}dt^2-e^{-2\phi/(d-2)}\gamma_{ij}dx^idx^j,
\end{equation}
where $i,j=1,\ldots,d$. So our fields are the scalar $\phi$ and the reduced spatial metric $\gamma_{ij}$. For the latter, we denote $\gamma=\det(\gamma_{ij})$, $R[\gamma]$ as its Ricci scalar, and $\Gamma[\gamma]^i{}_{jk}$ as its Christoffel symbols. The dimensional extension of Newton's constant is $G_d=G(\sqrt{4\pi e^{\gamma_E}}R_0)^{d-3}$, with $\gamma_E$ the Euler--Mascheroni constant and $R_0$ an arbitrary length scale. The harmonic-gauge residual $\mathcal C^i$ is
\begin{equation}
 {\cal C}^i=\gamma^{jk}
 \bigl(\Gamma[\gamma]^i{}_{jk}-\Gamma[\mathrm{flat}]^i{}_{jk}\bigr).
\end{equation}
In Cartesian coordinates, the flat connection vanishes. Keeping it makes ${\cal C}^i$ a genuine vector, which is required for our use of spherical coordinates below. The Einstein--Hilbert action $S_{\rm bulk}$, with harmonic gauge fixing, and its static value per unit time ${\mathcal S}_{\rm bulk}$ are
\begin{equation}\label{eq:Sbulk}
S_{\rm bulk} = \int \dd t\, {\mathcal S}_{\rm bulk} , \qquad {\mathcal S}_{\rm bulk}= - \frac1{16\pi G_d}\int\dd^dx\sqrt\gamma
 \left[\frac{d-1}{d-2}\,\gamma^{ij}\partial_i\phi\partial_j\phi-R[\gamma]
       +\frac12\,\gamma_{ij}{\cal C}^i{\cal C}^j\right].
\end{equation}

Label the two non-spinning point particles by $a=1,2$, with masses $m_1=M$, $m_2=m$, fixed positions $\mathbf x_a$ and proper times $\tau_a$. Their total worldline action $S_{\rm pp}$ and static action per unit time $\mathcal S_{\rm pp}$ are
\begin{equation}\label{eq:Spp}
S_{\rm pp} =  -\sum_{a=1,2} m_a \int \dd\tau_a = \int \dd t \,{\mathcal S}_{\rm pp}, \qquad
{\mathcal S}_{\rm pp} = - \sum_{a=1,2} m_a\, \frac{\dd\tau_a}{\dd t} = - \sum_{a=1,2} m_a\, e^{\phi(\mathbf x_a)},
\end{equation}
where we have restricted to particles at rest.

\subsection{Auxiliary sources}
Following ref.~\cite{Brunello}, rather than working directly with ${\mathcal S}_{\rm bulk}+{\mathcal S}_{\rm pp}$, whose couplings $e^{\phi(\mathbf x_a)}$ are nonlinear in the field, we first solve a simpler auxiliary problem. In it the scalar couples \emph{linearly} to two point sources of strengths $J_1,J_2$ at positions $\mathbf x_1,\mathbf x_2$, separated by $R=|\mathbf x_1-\mathbf x_2|$. Let $\operatorname{stat}_{\phi,\gamma_{ij}}$ mean evaluation on the solution of the field equations with these sources, constructed perturbatively in the sources with dimensional regularisation. We define the auxiliary stationary functional
\begin{equation}\label{eq:W}
W(J_1,J_2;R)=\operatorname{stat}_{\phi,\gamma_{ij}}\Bigl[{\mathcal S}_{\rm bulk}+J_1\,\phi(\mathbf x_1)+J_2\,\phi(\mathbf x_2)\Bigr].
\end{equation}
Since $e^{\phi}=1+\phi+\ldots$, the linear coupling reproduces the leading term $-m_a\phi(\mathbf x_a)$ of each particle action, with $J_a=-m_a$ at leading order. Below we show how the full exponential couplings are recovered from $W$ exactly.

Two properties make $W$ useful. First, due to stationarity in the fields, differentiating it with respect to a source acts only on the explicit source term:
\begin{equation}\label{eq:dW}
\frac{\partial W}{\partial J_a}=\phi(\mathbf x_a),
\end{equation}
the field at that source, evaluated with the same dimensional prescription as $W$. Second, evaluating an action on the solution of its sourced field equations sums all tree diagrams~\cite{BoulwareBrown}, so $W$ is the generating functional of classical (tree-level) connected correlators. Strictly speaking, $W$ is the generating functional per unit time, due to our static reduction. Expanding in the sources,
\begin{equation}\label{eq:Wexp}
W=\sum_{a,b\geq1}\frac{J_1^aJ_2^b}{a!\,b!}\,\Gamma_{a,b}(R),
\qquad \Gamma_{a,b}=\Gamma_{b,a}=\left(\frac GR\right)^{a+b-1}\overline\Gamma_{a,b}.
\end{equation}
The $\Gamma_{a,b}(R)$ are the connected static correlators of ref.~\cite{Brunello}, with $a$ insertions at $\mathbf x_1$ and $b$ at $\mathbf x_2$. The second equality, and the statement that the dimensionless coefficients $\overline\Gamma_{a,b}$ are pure numbers, refer to the dimensionally regulated physical limit; at generic $d$, the $\Gamma_{a,b}$ retain their dependence on $G_d$ and on the dimension. Terms with only one source are scaleless self-energies, and vanish in dimensional regularisation, which is why the sum starts at $a,b=1$. Moreover, the simultaneous transformation $\phi\to-\phi$, $J_a\to-J_a$ leaves \eqref{eq:W} invariant; we will refer to this as scalar sign reversal. Consequently, $W(-J_1,-J_2;R)=W(J_1,J_2;R)$, so every coefficient with $a+b$ odd vanishes.

\subsection{Conversion to the physical potential}
Here, we relate the auxiliary functional $W$ to the static energy, and then to the static potential. For a static configuration, the action per unit time is minus the energy. With the worldline coupling~\eqref{eq:Spp}, the total static energy is therefore
\[
 E_{\rm static}=-\operatorname{stat}_{\phi,\gamma_{ij}}\Bigl[{\mathcal S}_{\rm bulk}-\sum_{a=1,2}m_ae^{\phi(\mathbf x_a)}\Bigr].
\]
The physical coupling to the particles is nonlinear in $\phi$. To make it linear, we introduce two independent auxiliary field values $\Phi_1,\Phi_2$, standing for $\phi(\mathbf x_a)$, with the sources $J_a$ as Lagrange multipliers:
\[
 E_{\rm static}=-\operatorname{stat}_{\phi,\gamma_{ij},J_a,\Phi_a}\Bigl[{\mathcal S}_{\rm bulk}+\sum_{a=1,2}J_a\bigl(\phi(\mathbf x_a)-\Phi_a\bigr)-\sum_{a=1,2}m_ae^{\Phi_a}\Bigr],
\]
where $\operatorname{stat}$ denotes evaluation at a simultaneous stationary point in all the variables listed.
Working perturbatively on the branch connected to the free theory, we may evaluate this stationary value by eliminating either the auxiliary variables or the gravitational fields first. Eliminating $J_a$ and $\Phi_a$ first, stationarity in $J_a$ imposes $\Phi_a=\phi(\mathbf x_a)$, and we recover the previous expression. Eliminating instead the gravitational fields first, at fixed sources, leads by~\eqref{eq:W} to
\begin{equation}\label{eq:transform}
 E_{\rm static}=-\operatorname{stat}_{J_a,\Phi_a}
 \left[W(J_1,J_2;R)-\sum_{a=1,2}J_a\Phi_a-\sum_{a=1,2}m_ae^{\Phi_a}\right].
\end{equation}
Varying $J_a$ and $\Phi_a$ gives, respectively,
\begin{equation}
 \Phi_a=\frac{\partial W}{\partial J_a},\qquad J_a=-m_ae^{\Phi_a}.
\end{equation}
These equations determine stationary sources $J_a^\star(M,m,R)$ such that $J_a^\star\to-m_a$ as $G\to0$. Eliminating the $\Phi_a$ and subtracting the rest energies of the two bodies gives the static potential,
\begin{equation}\label{eq:Vexact}
 V_{\rm static}= E_{\rm static}-M-m =
 \left[-W+\sum_{a=1,2}J_a\,\frac{\partial W}{\partial J_a}-\sum_{a=1,2}J_a\right]_{J_a=J_a^\star}-M-m.
\end{equation}

\subsection{The response at first self-force order}
At fixed $J_1,R$, define $w_1=\partial_{J_2}W|_{J_2=0}$ and $w_2=\partial_{J_2}^2W|_{J_2=0}$. Since the one-source part of the stationary functional vanishes, the expansion in the light source is
\begin{equation}\label{eq:Wlight}
 W=J_2w_1(J_1;R)+\frac12J_2^2w_2(J_1;R)+O(J_2^3).
\end{equation}
According to \eqref{eq:dW}, the first function, $w_1$, is the field produced by the heavy source alone, evaluated at the light body's position. The second, $w_2$, measures how that field changes when a small light source is added: it is the field produced at the light body by its own source in the presence of the heavy body, and it includes the change in the spatial metric, which feeds back on the scalar.

We now expand the potential~\eqref{eq:Vexact} in the physical light mass $m$, keeping $M$ and $R$ fixed. At $m=0$ we have $J_2^\star=0$, so differentiating~\eqref{eq:Wlight} gives $\Phi_1=0$, the regularised self-field at the heavy source, and hence $J_1^\star=-M$. When we turn on $m$, the condition $J_2^\star=-me^{\Phi_2}$ already contains a factor of $m$, so the zeroth-order field $\Phi_2=w_1(-M;R)$ at the light position is enough:
\begin{equation}
J_2^\star=-m\,e^{w_1(-M;R)}+O(m^2).
\end{equation}
In the following, $w_1$, $w_2$ and their derivatives with respect to $J_1$ are understood to be evaluated at $J_1=-M$, unless their arguments are displayed explicitly. The light body produces the field $\Phi_1=J_2^\star\,\partial_{J_1}w_1+O(m^2)$ at the heavy body, so the condition $J_1=-Me^{\Phi_1}$ shifts the heavy auxiliary source:
\begin{equation}
 J_1^\star=M(-1+m\,e^{w_1}\partial_{J_1}w_1)+O(m^2).
\end{equation}

Finally, the expression in~\eqref{eq:transform} is stationary in $J_a$ and $\Phi_a$, so the derivative of $E_{\rm static}$ with respect to $m$ equals the derivative of its explicit dependence on $m$, which enters only through $-me^{\Phi_2}$. Subtracting the rest mass as in~\eqref{eq:Vexact},
\begin{equation}
\label{eq:dVdm}
 \frac{dV_{\rm static}}{dm}=e^{\Phi_2}-1 .
\end{equation}
At first order in $m$, the field at the light body changes for two reasons: the heavy source has shifted, and the light source adds its own field. Expanding $\Phi_2=\partial W/\partial{J_2}=w_1(J_1^\star)+J_2^\star w_2+O(m^2)$ about $J_1=-M$ gives
\begin{equation}
 \Phi_2=w_1+m\,e^{w_1}\big[M(\partial_{J_1}w_1)^2-w_2\big]+O(m^2).
\end{equation}
Exponentiating as in \eqref{eq:dVdm}, and integrating from $m=0$, we obtain
\begin{equation}\label{eq:Vw2}
 V_{\rm static}=m(e^{w_1}-1)
 +\frac{m^2e^{2w_1}}2\left[M(\partial_{J_1}w_1)^2-w_2\right]+O(m^3).
\end{equation}

Let us now determine $w_1$ and $\partial_{J_1}w_1$, evaluated at $J_1=-M$ as needed above. For the heavy body alone, $w_1$ is the scalar field evaluated at $r=R$. We may use its physical $d=3$ expression here, because of the regularity as $d\to3$, which follows from the recurrence in
Appendix~\ref{sec:app_radial}. In this limit, the background is the
four-dimensional Schwarzschild solution. In harmonic coordinates,
$\bar\phi(r)=-\atanh(GM/r)$, so that
$e^{2\bar\phi}=(r-GM)/(r+GM)$. Hence, in the one-source problem for which $J_1=-M$, we have
\begin{equation}
 w_1(J_1;R)
 =\atanh\frac{GJ_1}{R}.
\end{equation}
Differentiating at fixed
$R$ and then evaluating at $J_1=-M$, we obtain
\begin{equation}\label{eq:w1}
 w_1=-\atanh u,\qquad
 \partial_{J_1}w_1=\frac{G/R}{1-u^2},\qquad
 u=\frac{GM}{R}.
\end{equation}
In particular, $e^{w_1}=\sqrt{(1-u)/(1+u)}$ is the redshift factor
at the light body, so the term linear in $m$ in~\eqref{eq:Vw2}
corresponds to the probe potential~\eqref{eq:v0SF}. 
Comparing~\eqref{eq:w1} with~\eqref{eq:Wexp}, the zeroth-self-force correlators are $\overline\Gamma_{1,n}=\overline\Gamma_{n,1}=(n-1)!$ for odd $n$, and zero for even $n$, in agreement with ref.~\cite{Brunello}.

What remains is to compute the response $w_2$. This is the goal for the remainder of the paper. We can define the dimensionless response,
\begin{equation}\label{eq:Cdef}
 C(u)=\frac RG\, w_2(-M;R)
 =\sum_{n\geq1}\frac{\overline\Gamma_{2,2n}}{(2n)!}u^{2n},
\end{equation}
where the second equality follows from \eqref{eq:Wexp}. Comparing~\eqref{eq:Vw2} with~\eqref{eq:Vsf} gives the desired first self-force contribution,
\begin{equation}\label{eq:v1fromC}
{v_{1SF}(u)=\frac{e^{-2\atanh u}}2
 \left[\frac{u^2}{(1-u^2)^2}-uC(u)\right].}
\end{equation}
Sections~\ref{sec:response}--\ref{sec:summation} compute the response $w_2$, or equivalently $C$, using dimensional regularisation of the specified point-source action, with scaleless integrals set to zero and no additional finite terms introduced on the particle worldlines (i.e.~no finite-size or other local operators).

We note that the potential is defined in harmonic gauge here. It is coordinate dependent and is not by itself a gauge-invariant observable of the binary.

\section{The response problem}\label{sec:response}
In this section, we set up the calculation of the response $w_2$ in terms of a Green's function in general dimension, fixed by matching to the point-source EFT.

\subsection{The background and the fixed heavy source}\label{sec:background}
It is useful to write $p=d-2$ and $N=p+1$, so that $p$ is the exponent of the Newtonian falloff $r^{-p}$, and $N$ is the dimension of the angular sphere; the physical case is $p=1$. With the heavy source at $\mathbf x_1=0$ and the light source at $\mathbf x_2$, we have $|\mathbf x_2|=R$ and use $r=|\mathbf x|$ for the radial coordinate. For the field equations, we vary $\mathcal S_{\rm bulk}$.
A bar on a quantity will denote its (heavy-source) background value, and $\delta$ will denote its first variation; e.g.~$\phi=\bar\phi+\delta\phi$. On the unit $N$-sphere, let $\theta^A$ be angular coordinates ($A,B=1,\ldots,N$), $\Omega_{AB}$ the metric and $D_A$ its covariant derivative, with $\dd\Omega_N^2=\Omega_{AB}\dd\theta^A\dd\theta^B$. The sphere has area $\Omega_N=2\pi^{(N+1)/2}/\Gamma((N+1)/2)$. We parametrise the background by
\[
 \mu=\frac{8\pi G_dM}{N\Omega_N},
\]
which reduces to $GM$ in $d=3$. Spherical symmetry then allows us to write the reduced spatial metric in harmonic radius as
\[
 \bar\gamma_{ij}\dd x^i\dd x^j=A(r)\dd r^2+B(r)\dd\Omega_N^2.
\]

Holding the heavy source fixed has a useful interpretation as a
Gauss-law condition. Stationarity with respect to the scalar in \eqref{eq:W} implies
\begin{equation}\label{eq:scalar_source}
 \partial_i\!\left(\sqrt\gamma\,\gamma^{ij}\partial_j\phi\right)
 =-\frac{8\pi p G_d}{N}
   \sum_{a=1,2}J_a\,\delta^{(d)}(\mathbf x-\mathbf x_a).
\end{equation}
For the heavy-source background, $J_2=0$. Away from the heavy source,
the equation for the background scalar $\bar\phi$ is
\[
 \frac{\dd}{\dd r}
 \left(S_r\,\bar\phi'\right)=0, \qquad \text{where} \quad S_r=\frac{B^{N/2}}{\sqrt A},
\]
and where primes denote $r$ derivatives. So the scalar flux $\Omega_NS_r\bar\phi'$ through a sphere around the heavy source is constant, and integrating~\eqref{eq:scalar_source} over the enclosed ball fixes it:
\[
 \Omega_N S_r\bar\phi'
 =-\frac{8\pi G_dp}{N}J_1,
 \qquad\text{so, for $J_1=-M$},\quad
 S_r\bar\phi'=p\mu.
\]
When we add the light source, the scalar field and the spatial
metric change. However, the integrated flux through any surface
enclosing only the heavy source remains fixed, because the enclosed
source strength $J_1$ is unchanged. This is the condition appropriate
to the calculation of $w_2$, which holds the auxiliary source fixed.

For later reference, at $p=1$ the background is
\begin{equation}\label{eq:background}
 \bar\phi=-\atanh(\mu/r),\qquad A=1,\qquad
 B=r^2-\mu^2,\qquad \mu=GM.
\end{equation}
Hence, $e^{2\bar\phi}=(r-\mu)/(r+\mu)$, and the areal radius is $r_{\mathrm a}=r+\mu$.

\subsection{Coupled scalar and metric perturbations}\label{sec:perturbations}

The response $w_2$ is the connected static two-point function of
scalar perturbations on the heavy-source background. Recall, from \eqref{eq:Wlight},
\begin{equation}
 w_2(J_1;R)
 = \left.\frac{\partial^2 W}{\partial J_2^2}
   \right|_{J_2=0},
 \qquad J_1\ \text{fixed}.
\end{equation}
The two derivatives with
respect to $J_2$ insert two scalar fields at the light-body position,
while keeping $J_1$ unexpanded resums all heavy-source insertions
into the background. We are therefore interested in the regularised coincident limit (i.e.~same position $\mathbf x_2$) of a Green's function, which is the inverse of the operator governing the quadratic-in-perturbations part of the action. Because scalar $\phi$ and metric $\gamma_{ij}$ perturbations mix, we need the scalar--scalar component of the inverse of the full coupled
operator. 
We will now implement this construction by decomposing
the perturbations into angular harmonics.

Spherical symmetry allows us to treat each angular multipole separately. Let $\ell=0,1,2,\ldots$ be its degree, $\lambda=\ell(\ell+p)$ the Laplacian eigenvalue, and $\dd\Omega$ the unit-sphere measure. We choose a real harmonic $Y$, normalised by $\int Y^2\,\dd\Omega=1$, with $D^AD_AY=-\lambda Y$, and angular indices raised using $\Omega^{AB}$. Its trace-free tensor is $Y_{AB}=D_AD_BY+\lambda\Omega_{AB}Y/N$. The scalar source then excites five radial perturbation functions $\psi,f,v,c,e$:
\begin{align}\label{eq:fields}
 \delta\phi&=\psi Y,& \delta\gamma_{rr}&=AfY,\notag\\
 \delta\gamma_{rA}&=rvD_AY,&
 \delta\gamma_{AB}&=B(c\Omega_{AB}Y+eY_{AB}).
\end{align}
Although we ultimately want only $\psi$, its effect on the metric feeds back into the scalar equation, so we have to invert the {coupled} system. For $\ell=0$ the vector and tensor perturbations are absent, while for $\ell=1$ the tensor perturbation is absent. We omit these absent components when forming the corresponding operators.

Let us write $q=(\psi,f,v,c,e)^T$ for the column of radial functions, with $T$ denoting transpose and absent components understood to be omitted. In the full harmonic expansion, denote this column by $q_{\ell a}$, where $a$ labels the real orthonormal harmonics of degree $\ell$. Let ${\mathcal S}^{(2)}_{\rm bulk}$ denote the quadratic contribution of these perturbations to the static bulk action per unit time, summed over all harmonics. Since different harmonics do not mix, angular integration and integration by parts define the radial Hessians $H_\ell$ through
\begin{equation}
 16\pi G_d\,{\mathcal S}^{(2)}_{\rm bulk}
 =-\frac12\sum_{\ell=0}^{\infty}\sum_a
 \int_0^\infty q_{\ell a}^T H_\ell q_{\ell a}\,\dd r,
\end{equation}
where spherical symmetry makes $H_\ell$ independent of $a$, and the minus is for convenience. For notational simplicity, we suppress the harmonic labels in the following single-mode equations, writing $q=q_{\ell a}$ and $H=H_\ell$. The factor $16\pi G_d$ removes the overall coupling. For $\ell\geq2$, $H$ is a $5\times5$ matrix of second-order differential operators in $r$, depending on the background and on $\ell,p$; the monopole and dipole have $3\times3$ and $4\times4$ matrices, respectively.

The light source couples only to the scalar, through $J_2\phi(\mathbf x_2)$ in~\eqref{eq:W}. If $\hat{\mathbf x}_2=\mathbf x_2/R$ denotes its direction, the perturbation of this coupling in our harmonic is $J_2Y(\hat{\mathbf x}_2)\psi(R)$.    Varying ${\mathcal S}^{(2)}_{\rm bulk}$ together with this coupling gives
\begin{equation}\label{eq:sourced}
 Hq=16\pi G_dJ_2Y(\hat{\mathbf x}_2)\,\delta(r-R)\,\boldsymbol e_\psi,
\end{equation}
where $\boldsymbol e_\psi$ is the unit column selecting the scalar component. With $\Id$ denoting the identity, we define the matrix radial Green's function $\mathcal G$ by
\begin{equation}
 H_r\mathcal G(r,R;p)=\Id\,\delta(r-R),
\end{equation}
where $H_r$ is the full matrix operator $H$, acting on the first radial argument of $\mathcal G$. The perturbations then take the form
\begin{equation}
 q(r)=16\pi G_dJ_2Y(\hat{\mathbf x}_2)\,\mathcal G(r,R;p)\,\boldsymbol e_\psi.
\end{equation}
The scalar response per unit radial source is $16\pi G_d\mathcal G_{\psi\psi}(R,R;p)$.

We will need to sum over all harmonics. At integer spatial dimension, let $Y_{\ell a}$ be a real orthonormal basis, where $a$ labels the $(2\ell+p)\omega_\ell(p)$ harmonics of degree $\ell$ and 
\begin{equation}
\label{eq:omegal}
 \omega_\ell(p)=\frac{\Gamma(\ell+p)}{\Gamma(p+1)\Gamma(\ell+1)}. \qquad \text{Then} \qquad \sum_aY_{\ell a}(\hat{\mathbf x}_2)^2
 =\frac{(2\ell+p)\omega_\ell(p)}{\Omega_N}
\end{equation}
by the addition theorem.
We continue these angular factors in $d$ and write $[\cdots]_{\rm DR}$ for the dimensionally regulated sum, to be defined in Section~\ref{sec:angular}. The response is
\begin{equation}\label{eq:modesum}
 w_2=16\pi G_d\left[\sum_{\ell\geq0}
 \frac{(2\ell+p)\omega_\ell(p)}{\Omega_N}\,
 \mathcal G_{\psi\psi,\ell}(R,R;p)\right]_{\rm DR},
\end{equation}
where the subscript $\ell$ identifies the harmonic sector. At $p=1$, where $\Omega_2=4\pi$ and $\omega_\ell=1$, this becomes $w_2=4G[\sum_{\ell\geq0}(2\ell+1)\mathcal G_{\psi\psi,\ell}(R,R;1)]_{\rm DR}$.

There are two distinct issues with this coincident evaluation. For each fixed $\ell$, the radial Green's function is continuous at $r=R$ and only its first derivative has a source jump, but we still have to show that its dimensional continuation has no pole at $p=1$. We do this in Sections~\ref{sec:continuation} and~\ref{sec:exceptional}. Even when these individual radial values are finite, their infinite angular sum needs regularisation.

It is convenient to normalise each radial contribution by its flat-space value.
When the heavy source is absent, the scalar and metric perturbations
decouple, and the scalar radial Green's function at equal arguments is
\[
 \mathcal G^{\mathrm{flat}}_{\psi\psi,\ell}(R,R;p)
 =\frac{p}{2N(2\ell+p)R^p}.
\]
With $z_R=\mu/R^p$ measuring the background strength at the source,
define
\begin{equation}\label{eq:Tdef}
 T_\ell(p,z_R)
 =\frac{\mathcal G_{\psi\psi,\ell}(R,R;p)}
        {\mathcal G^{\mathrm{flat}}_{\psi\psi,\ell}(R,R;p)}
 =\frac{2N}{p}(2\ell+p)R^p
   \mathcal G_{\psi\psi,\ell}(R,R;p).
\end{equation}
By construction, $T_\ell(p,0)=1$. In terms of these normalised
radial contributions, the response in the physical dimension is
\begin{equation}\label{eq:CT}
 C(u)=\left[\sum_{\ell\geq0}T_\ell(1,u)\right]_{\rm DR}.
\end{equation}
Although each flat-space contribution equals one, their
dimensionally regulated sum vanishes, consistently with the
vanishing flat-space self-field. The prescription for the angular
sum is explained in Section~\ref{sec:angular}.

The gauge-fixing term in~\eqref{eq:Sbulk} has a consequence that we must keep track of.
In the stationary problem defining $W$ in~\eqref{eq:W}, the
source positions are held fixed, so we do not impose their
equations of motion. Their forces instead enter an equation
for the gauge residual. In Cartesian coordinates, with
$\mathcal C_k=\gamma_{ki}\mathcal C^i$, the sourced field equations
imply the exact Ward identity
\begin{equation}\label{eq:ward}
 \partial_i\partial_j
 \left(\sqrt\gamma\,\gamma^{ij}\mathcal C_k\right)
 =
 16\pi G_d\sum_{a=1,2}
 J_a\,\partial_k\phi_{\rm reg}(\mathbf x_a)\,
 \delta^{(d)}(\mathbf x-\mathbf x_a),
\end{equation}
where $J_a\partial_k\phi_{\rm reg}(\mathbf x_a)=\partial W/\partial x_a^k$ is the
force on source $a$, and the spatial derivative is taken before evaluating at the
source, using the same dimensional prescription as $W$.
At the stationary point, the first variation of the action vanishes under an
infinitesimal spatial diffeomorphism acting on the fields alone, with the
source positions and the flat reference metric held fixed. The invariant
part of the action changes only by a boundary term, the source couplings
give the forces, and integrating the variation of the gauge-fixing term
by parts twice gives the left-hand side of~\eqref{eq:ward}.
Nonzero forces therefore require a nonzero gauge residual,
which we retain when inverting $H$. Conversely, if both forces
vanished, the same point-source prescription would give $\mathcal C^i=0$,
so harmonic gauge would hold exactly.

Since the heavy background is harmonic,
$\partial_j(\sqrt{\bar\gamma}\,\bar\gamma^{ij})=0$.
At linear order, the left-hand side of~\eqref{eq:ward} becomes
\[
 \partial_i\left(
 \sqrt{\bar\gamma}\,\bar\gamma^{ij}
 \partial_j\delta\mathcal C_k\right).
\]
Each Cartesian component of the residual thus obeys a scalar
equation on the fixed background. 
Decomposing this equation into angular harmonics gives the
first two radial equations in~\eqref{eq:reduced}, as discussed
in Section~\ref{sec:decoupled}.
The force on the heavy source is included in
Section~\ref{sec:dipole}.

\subsection{The point-source prescription for the inverse} \label{sec:prescription}
The equation $H\mathcal G=\Id\delta$ fixes the jump at the light source but leaves freedom to add homogeneous solutions. To fix that freedom, we return to the perturbative definition of the EFT: start with the flat propagator and insert the heavy background order by order, at generic dimension. This determines the inner ($r<R$) and outer ($r>R$) radial branches of $\mathcal G$ without a separate horizon boundary condition.

To start the background expansion, we need the flat-space radial
solutions. We use $j$ to denote their radial powers, while $\ell$
continues to label the angular multipole. For fixed $\ell\geq2$,
the flat equations have five independent solutions regular at
the origin, each of the form
\begin{equation}
 q(r)=r^j\boldsymbol v,
\end{equation}
where the constant column $\boldsymbol v$ specifies the relative
values of the five perturbation functions. Substituting this form
into the flat equations gives
\begin{equation}
\label{eq:flatpowers}
 j=\ell-2,\quad \ell,\quad \ell,\quad \ell,\quad \ell+2.
\end{equation}
In particular, three independent solutions have the same radial power
$r^\ell$, but different columns $\boldsymbol v$. The flat matrix
used to obtain these powers is given in
Appendix~\ref{sec:app_radial}. For each of these solutions, there is a corresponding solution
proportional to $r^{-j-p}$, which decays at infinity. The flat
Green's function uses the $r^j$ solutions on the inner side
$r<R$ and the $r^{-j-p}$ solutions on the outer side $r>R$.
We then construct their corrections order by order in the heavy
background, and call the resulting solutions \emph{inner} and \emph{outer}, respectively.

For the monopole and dipole, fewer perturbation functions are
present. The corresponding lists of inner radial powers are
$j=0,0,2$ for $\ell=0$ and $j=1,1,1,3$ for $\ell=1$,
again with an outer solution proportional to $r^{-j-p}$ for
each entry.

The flat propagator fixes the homogeneous terms as well as solving
the differential equation. To see this explicitly, let $\rho$ denote
the integration radius. In each degree-$j$ block of the flat system,
the radial integration reduces to the Green's function of the operator
\[
 -\frac{\dd}{\dd r}\left(r^{p+1}\frac{\dd}{\dd r}\right)
 +j(j+p)r^{p-1}.
\]
Its Green's function is
\begin{equation}\label{eq:G0j}
 \mathcal G_{0,j}(r,\rho)
 =\frac{r_<^j r_>^{-j-p}}{2j+p}, \qquad \text{with}\quad r_<=\min(r,\rho),\quad r_>=\max(r,\rho).
\end{equation}
For example, the scalar component of the full flat inverse is
$p/(2N)$ times $\mathcal G_{0,\ell}$, consistently with the
flat-space value used in~\eqref{eq:Tdef}.
For an exponent $b$, applying $\mathcal G_{0,j}$ to the
power-law source $\rho^{b-2}$ with radial measure
$\rho^{p+1}\dd\rho$ gives
\begin{align}
 \int_0^\infty \mathcal G_{0,j}(r,\rho)\rho^{b+p-1}\,\dd\rho
 &=
 \frac{1}{2j+p}\left[
 r^{-j-p}\int_0^r \rho^{b+j+p-1}\,\dd\rho
 +r^j\int_r^\infty \rho^{b-j-1}\,\dd\rho
 \right]\notag\\
 &=-\frac{r^b}{(b-j)(b+j+p)}.
 \label{eq:intG0}
\end{align}
Solving the differential equation alone would allow arbitrary
multiples of $r^j$ and $r^{-j-p}$ to be added, whereas the integral
gives a definite answer. We first evaluate it where both integrals
converge, then define its value elsewhere by meromorphic continuation
in $b$ and $p$. This fixes the homogeneous additions for generic
exponents. The cases that remain resonant at generic $p$ require
the source information discussed below. We use this prescription at
every step of the Born series,
which inserts the heavy background repeatedly into the flat
propagator. Whenever derivatives act on a propagator, we retain
the delta functions arising from its derivative jump at the source.

For calculation, we organise these insertions as a recurrence. Each power of $\mu$ lowers the radial exponent by $p$ (recall $\mu/r^p$ is dimensionless), so for a trial exponent $\alpha$ and a constant column $\boldsymbol v$ we define matrices $\mathsf H_k(\alpha,p)$ by
\begin{equation}\label{eq:powerH}
 H(r^\alpha\boldsymbol v)=r^{\alpha+p-1}\sum_{k\geq0}
             (\mu/r^p)^k\mathsf H_k(\alpha,p)\boldsymbol v,
\end{equation}
where $k\geq0$ counts powers of $\mu$.
We choose one of the flat radial exponents, $\beta=j$ or
$\beta=-j-p$, and a corresponding constant column
$\boldsymbol v_0$ satisfying
$\mathsf H_0(\beta,p)\boldsymbol v_0=0$.
We seek its background corrections in the form
 \begin{equation}
q=r^\beta\sum_{k\geq0}(\mu/r^p)^k\boldsymbol v_k.
\end{equation}
This is a Frobenius expansion, i.e.~a fixed leading radial power
followed by its weak-field corrections. For positive $k$, with $i$
a summation index, substitution into $Hq=0$ gives
\begin{equation}\label{eq:recurrence}
 \mathsf H_0(\beta-kp,p)\boldsymbol v_k
 =-\sum_{i=1}^k\mathsf H_i(\beta-(k-i)p,p)\boldsymbol v_{k-i}.
\end{equation}
For example,
$\mathsf H_0(\beta-p,p)\boldsymbol v_1
=-\mathsf H_1(\beta,p)\boldsymbol v_0$
is an equation for the first correction, driven by the known
zeroth-order solution. We solve it by applying the flat propagator
to the source on the right. Equation~\eqref{eq:intG0} specifies the
radial integration and its meromorphic continuation. When the matrix
on the left becomes singular only at isolated dimensions, we first
perform the inversion at generic $p$ and then continue the result.
The matrices $\mathsf H_0$ and $\mathsf H_1$, and the algorithm generating the higher ones from the action~\eqref{eq:Sbulk}, are given in Appendix~\ref{sec:app_radial}; the ancillary code generates them through sixth order.

The flat solutions with a constant Cartesian metric perturbation
have leading degree $j=0$. Their first correction has exponent $-p$,
which is also the decaying degree-zero exponent for every $p$.
Thus $\mathsf H_0(-p,p)$ is singular even at generic dimension, and
the vacuum recurrence alone does not determine all coefficients.
To see why the missing information is at the source, let $\nabla^2$
denote the flat Laplacian and recall
$\nabla^2r^{-p}=-p\Omega_N\delta^{(d)}(\mathbf x)$.
Changing the coefficient of $r^{-p}$ changes the point source,
although it makes no difference to the vacuum equation away from
the origin.

We can recover this information by placing the heavy source in a
constant infinitesimal metric perturbation $h_{ij}$. Write
$n^i=x^i/r$ and $\gamma_{ij}=\delta_{ij}+h_{ij}$ at leading order
in $\mu$, with $\delta_{ij}$ the Euclidean metric. The leading-order
scalar field and its variation are
\begin{equation}\label{eq:contact}
 \phi=-\mu(\gamma_{ij}x^ix^j)^{-p/2},\qquad
 \delta\phi=\frac{p\mu}{2}r^{-p}h_{ij}n^in^j.
\end{equation}
The trace of $h_{ij}$ contributes to the monopole, while its
traceless part contributes to the quadrupole. In the monopole,
\eqref{eq:contact} fixes the scalar coefficient by preserving the
heavy-source flux. In the quadrupole, it agrees with the scalar
coefficient already fixed by the vacuum recurrence. In both sectors, the construction~\eqref{eq:contact} has no order-$\mu$ metric correction,
fixing the remaining metric freedom to zero.

Within the point-source prescription~\eqref{eq:intG0}, dimensional scaling excludes further contact-term ambiguities at generic
dimension.\footnote{
Before angular reduction, the coefficient multiplying $n$ spatial
derivatives at order $\mu^k$ has radial degree $n-2-kp$.
The spatial delta function has degree $-p-2$, so a contact term
with this scaling would require $(k-1)p-n$ derivatives of the
delta function. For fixed integers $k\geq1$ and $n\geq0$, the
required derivative count is independent of $p$ only if $k=1$,
and is then nonnegative only if $n=0$. This leaves the contact term~\eqref{eq:contact}. Resonances occurring only at special dimensions
are treated by meromorphic continuation.
}

\subsection{Normalisation and control of dimensional poles}
We construct the Green's function by joining inner and outer
solutions at the light source, then bound its possible pole order
at $p=1$.

With radial coefficient matrices $\mathsf K,\mathsf S,\mathsf U$
and primes denoting $r$ derivatives, the Hessian takes the form
\begin{equation}\label{eq:Hform}
 Hq=-(\mathsf Kq')'+\mathsf Sq'+(\mathsf U+\mathsf S'/2)q,\qquad
 \mathsf K^T=\mathsf K,\quad
 \mathsf U^T=\mathsf U,\quad
 \mathsf S^T=-\mathsf S.
\end{equation}
For radial columns $a,b$, define the pairing
\begin{equation}\label{eq:pair}
 \B(a,b)=a^T\mathsf Kb'-a'^T\mathsf Kb-a^T\mathsf Sb,\qquad
 \B'=(Ha)^Tb-a^THb.
\end{equation}
The pairing is antisymmetric and, by the second relation, constant when $Ha=Hb=0$.
If we had a single equation, this would reduce to the weighted Wronskian
$\mathsf K(ab'-a'b)$. Similarly to how the Wronskian is used to normalise a Green's function at the source, in our case we will have a matrix-valued Wronskian for a matrix-valued Green's function.

Let $U_{\rm in}$ and $U_{\rm out}$ collect the independent inner and outer solutions of Section~\ref{sec:prescription} as columns, each column containing all radial perturbation functions of one solution. For $\ell\geq2$ there are five columns in each matrix, labelled by the exponents~\eqref{eq:flatpowers}; the monopole and dipole have three and four. For matrix arguments, $\B$ denotes the matrix of all column pairings.

An inner column labelled by $i$ starts as $r^i$, and its
outer partner starts as $r^{-i-p}$. The radial weights of
$\mathsf K$ and $\mathsf S$ are $r^{p+1}$ and $r^p$, respectively,
multiplying the series in $\mu/r^p$ due to the background. So the order-$\mu^k$
contribution to a pairing of columns labelled by $i,j$ has
radial dependence
\[
 \mu^k
 \begin{cases}
 r^{i+j+(1-k)p}, & \text{inner--inner},\\
 r^{-i-j-(k+1)p}, & \text{outer--outer},\\
 r^{i-j-kp}, & \text{inner--outer}.
 \end{cases}
\]
Conservation of the pairing requires its coefficient to vanish unless the
exponent is zero.

At generic $p$, the inner--inner exponent vanishes only
for $i=j=0$, $k=1$. In the quadrupole there is one such column,
whose self-pairing vanishes by antisymmetry. In the monopole, the only potentially nonzero pairing is
between the constant scalar perturbation ($\psi=\mathrm{constant}$,
$f=c=0$) and the solution generated by a radial rescaling.
This pairing measures the variation of the heavy-source flux,
which the source condition~\eqref{eq:contact} sets to zero. The outer--outer exponent never
vanishes. Therefore, we have the pairing identities
\[
 \B(U_{\rm in},U_{\rm in})=0,\qquad
 \B(U_{\rm out},U_{\rm out})=0.
\]
The inner--outer exponent vanishes only for $k=0$, $i=j$,
so the cross-pairing has exactly its flat-space value (order $\mu^0$):
\[
 \mathcal W\equiv\B(U_{\rm in},U_{\rm out})
 =\left[
 U_{\rm in}^T\mathsf K U_{\rm out}'
 -(U_{\rm in}')^T\mathsf K U_{\rm out}
 \right]_{\mu=0}.
\]
This is the Wronskian matrix. All matrices here are evaluated at the same radius, and
$\mathsf S=0$ in flat space. The flat solution bases and
$\mathsf K$ are analytic near $p=1$, with $\mathsf K$ nonsingular
(Appendix~\ref{sec:app_radial}). A null vector of $\mathcal W$
would give a nonzero outer solution pairing to zero with
every inner and outer solution. Since these form a complete
basis, pairing with arbitrary initial values and derivatives
would force that solution to vanish. Thus $\mathcal W$ is
nonsingular, and $\mathcal W^{-1}$ is regular near $p=1$.

Pairing the sourced equation with the inner and outer
solutions and integrating across $r=R$ gives
\begin{equation}\label{eq:Green}
 \mathcal G(r,R)=
 \begin{cases}
 -U_{\rm in}(r)\mathcal W^{-T}U_{\rm out}(R)^T,&r<R,\\
 -U_{\rm out}(r)\mathcal W^{-1}U_{\rm in}(R)^T,&r>R,
 \end{cases}
\end{equation}
where $\mathcal W^{-T}=(\mathcal W^{-1})^T$.
The pairing identities ensure continuity and the unit source jump:
 \begin{equation}
 \mathcal G(R^+,R)=\mathcal G(R^-,R), \qquad
 -\mathsf K(R)\bigl[
 \partial_r\mathcal G(R^+,R)-\partial_r\mathcal G(R^-,R)
 \bigr]=\Id,
 \end{equation}
with $R^\pm$ denoting approach from above and below.
Exchanging $r$ and $R$ transposes~\eqref{eq:Green}, giving
$\mathcal G_{\psi\psi}(r,R)=\mathcal G_{\psi\psi}(R,r)$.

Possible dimensional poles remain in the solution columns.
To establish regularity of $\mathcal G_{\psi\psi}$,
Sections~\ref{sec:continuation} and~\ref{sec:exceptional} start by eliminating the most
singular coefficient of the coupled response, and then repeat for
the remaining poles. In the remainder of this section, we bound their order uniformly
throughout the weak-field expansion.

The background coefficients are regular at $p=1$, as follows from the recursion in
Appendix~\ref{sec:app_radial}. Poles can therefore enter through
the inversion of $\mathsf H_0(\beta-kp,p)$
in~\eqref{eq:recurrence}. On a flat sector labelled by $j$,
the inverse contains the factor
\[
 \frac{1}{(\beta-kp-j)(\beta-kp+j+p)},
\]
as in~\eqref{eq:intG0}, multiplying matrix factors analytic
near $p=1$. Identically vanishing factors correspond to resonances holding at generic $p$, already fixed by~\eqref{eq:contact}. Otherwise, a factor vanishing at $p=1$ produces
at most a simple pole. Both factors cannot vanish there together,
since their difference is $2j+1\neq0$.

At $p=1$, each weak-field correction ($\mu/r^p$) lowers the radial exponent by one, and whenever the exponent reaches another flat exponent, the recurrence~\eqref{eq:recurrence} can raise the pole order by at most one, as explained above. For $\ell\geq2$, the distinct flat exponents are $\ell+2,\ell,\ell-2$ and $-\ell+1,-\ell-1,-\ell-3$. An inner solution starts at one of the first three, so its descending exponent can meet at most the other five, while an outer solution starts among the last three and can meet at most two. Since each term in~\eqref{eq:Green} contains one inner and one outer solution, and $\mathcal W^{-1}$ is regular at $p=1$, the pole order of the Green's function is at most seven. For the monopole and dipole, whose distinct flat exponents at $p=1$ are $2,0,-1,-3$ and $3,1,-2,-4$, respectively, the same counting gives at most four. Once the descending exponent lies below every flat exponent, it meets no further ones, so these bounds hold at every weak-field order, and only finitely many negative Laurent powers need to be excluded.

\section{Dimensional continuation and the physical limit}\label{sec:continuation}

At $p=1$ ($d=3$), the radial equations reduce to Legendre-type
equations, which can be solved in closed form. Two issues require
us to retain dimensional continuation. First, it can leave a finite
constant, arising as a $0/0$ limit, that the three-dimensional equations alone do not fix (see~\eqref{eq:endconstant}). Second, the dimensionally continued
inverse may contain poles at $p=1$, which must be excluded before
identifying its finite part with a three-dimensional Green's
function. For example, if there were a simple pole,
\[
 \mathcal G(p)=\frac{\mathcal G_{-1}}{p-1}
              +\mathcal G_0+O(p-1),
\]
analyticity of $H(p)$ and the equation
$H(p)\mathcal G(p)=\Id\,\delta$ would give
\[
 H(1)\mathcal G_{-1}=0,\qquad
 H(1)\mathcal G_0+
 \left.\partial_pH\right|_{p=1}\mathcal G_{-1}
 =\Id\,\delta.
\]
The finite part $\mathcal G_0$ would therefore solve the
three-dimensional Green's-function equation only if the additional
term vanished. We address these issues by constructing one inner
solution at generic $p$, retaining the finite constant, and then
using it to exclude poles in the perturbation produced by the light
scalar source. Throughout this section $\ell\geq2$, and limits are
taken coefficient by coefficient in the weak-field expansion.

\subsection{Decoupled radial equations}\label{sec:decoupled}
We first look at the equations at $p=1$, before taking any limit of the dimensional Green's function, to see which parts of a perturbation they separate. It is convenient now to absorb $\mu$ into the variables, defining
\begin{equation}
\label{eq:xy}
x=r/\mu, \qquad y=R/\mu, \qquad \Delta=x^2-1, \qquad L_n=\partial_x(\Delta\partial_x)-n(n+1),
\end{equation}
where the latter is the Legendre operator of degree $n$. Primes now denote $x$ derivatives through Section~\ref{sec:summation}, unless explicitly stated otherwise. At $p=1$, the harmonic eigenvalue of a perturbation is $\lambda=\ell(\ell+1)$. The scalar equation with source $T$ reads
\[
L_\ell\psi+c'-f'/2+\lambda xv/\Delta=T(x),
\]
and its response to $T=\delta(x-y)$ defines $K_\ell(x,y)$. Since the scalar row of $H$ is $-4$ times the left-hand side and $\delta(r-R)=\delta(x-y)/\mu$, we have
\begin{equation}\label{eq:GK}
 \mathcal G_{\psi\psi}(r,R;1)=-\frac{K_\ell(x,y)}{4\mu}.
\end{equation}

We decompose the linearised harmonic-gauge residual as
\[
 \delta\mathcal C^r=\frac{w_r}{\mu}\,Y,\qquad
 \delta\mathcal C^A=\frac{w_\Omega}{\mu^2\Delta}D^AY,
\]
where $D^A=\Omega^{AB}D_B$. Equivalently, lowering the angular
index with the background reduced metric gives
$\bar\gamma_{AB}\delta\mathcal C^B=w_\Omega D_AY$.
In terms of the perturbations in~\eqref{eq:fields},
\[
 w_r=f'/2-c'+2x(f-c)/\Delta-\lambda xv/\Delta,\qquad
 w_\Omega=(xv)'+2v-f/2+(1-\lambda/2)e .
\]
We consider the combinations
\begin{equation}\label{eq:wpm}
 w_-=\frac{w_r+(\ell+1)w_\Omega/x}{2\ell+1},\qquad
 w_+=\frac{\ell w_\Omega/x-w_r}{2\ell+1},\qquad
 \Psi=\psi-\frac{xv}{\Delta}+\frac{e'}2+\frac{w_\Omega}{2x}.
\end{equation}
The five coupled linearised equations of motion imply three decoupled ones:
\begin{equation}\label{eq:reduced}
 L_{\ell-1}w_-=-\frac{4T}{(2\ell+1)\Delta},\qquad
 L_{\ell+1}w_+=\frac{4T}{(2\ell+1)\Delta},\qquad
 (L_\ell-4/\Delta)\Psi=T.
\end{equation}
The first two equations follow from the angular harmonic decomposition
of the linearised Ward identity~\eqref{eq:ward} and determine
the gauge residual produced by the source. The third equation governs the master variable $\Psi$.

The five coupled second-order equations have ten independent homogeneous solutions. The three equations in~\eqref{eq:reduced} account for six, two each for $w_-$, $w_+$ and $\Psi$. The remaining four are pure-gauge solutions that preserve the harmonic condition: they have $w_\pm=0$ and $\Psi=0$, and are generated by displacements whose Cartesian components solve the background scalar equation in degrees $\ell\pm1$, two solutions each. A homogeneous solution, such as the coefficient of a possible pole, is therefore determined by its gauge residual, its master variable and, if both vanish, its pure-gauge part. Section~\ref{sec:poles} examines a possible pole in exactly this order. The exceptional multipoles $\ell=0,1$ are treated in Section~\ref{sec:exceptional}.

\subsection{Analytic inner solution}\label{sec:inner}
Section~\ref{sec:poles} will show that the coupled perturbation
produced by the light scalar source has no pole at $p=1$.
That argument needs one known inner solution, analytic at $p=1$, with a nonzero master variable $\Psi$. We construct this inner solution at generic $p$ and denote
the resulting family by $q_{\rm phys}(r,p)$, normalised so that
$q_{\rm phys}(r,p)\sim r^\ell(1,0,0,0,0)^T$ at large $r$.
Its physical-dimensional limit $q_{\rm phys}(r,1)$ is given
in~\eqref{eq:physicalcolumn}.

Our strategy is to solve first the scalar equation on the fixed background (i.e.\ with the metric perturbations omitted), then to use its solution to build the coupled perturbation, and finally to add a pure-gauge perturbation that makes the
gauge residual vanish.
The important subtlety of the dimensional continuation is a finite constant that appears in the last step, through an integral that does not commute with the limit $p\to1$. Let us write $\AC\int$ for continued power integration without an added constant, so that a power $t^a$ integrates to $t^{a+1}/(a+1)$. Note that $\AC\int(p-1)t^{p-2}\dd t=t^{p-1}$ tends to $1$ as $p\to1$, although the integrand tends to zero. We will meet this type of contribution below.

We work at generic $p$, with $\lambda=\ell(\ell+p)$.
We temporarily use the areal radius $r_{\mathrm a}$, in which the background
is explicit at general dimension. The resulting perturbation will
later be expressed in the harmonic radius $r$ used for EFT matching.
In this coordinate the background fields are
\begin{equation}\label{eq:arealbackground}
 e^{2\bar\phi(r_{\mathrm a})}=1-\frac{2\mu}{r_{\mathrm a}^p},\qquad
 \bar\gamma=e^{2\bar\phi(r_{\mathrm a})/p}
 \left[e^{-2\bar\phi(r_{\mathrm a})}\,\dd r_{\mathrm a}^2+r_{\mathrm a}^2\dd\Omega_N^2\right].
\end{equation}
We call the linearised scalar equation with the metric
perturbations set to zero the \emph{frozen} equation.
We use $n$ for the angular degree of this fixed-background
scalar equation, since the construction requires its solutions
at degrees $\ell$, $\ell\pm1$ and $1$. Let $B_n(r_{\mathrm a})$
denote its growing radial solution, selected by the power-series
prescription and normalised so that
$B_n(r_{\mathrm a})\sim r_{\mathrm a}^n$ at infinity. It satisfies
\begin{equation}\label{eq:frozen}
 \partial_{r_{\mathrm a}}\bigl[(r_{\mathrm a}^{p+1}-2\mu r_{\mathrm a})\,\partial_{r_{\mathrm a}}B_n\bigr]
 -n(n+p)r_{\mathrm a}^{p-1}B_n=0.
\end{equation}
In the variable $X=r_{\mathrm a}^p/\mu-1$, setting $B_n=\mu^{n/p}F_{n/p}(X)$ turns this into a Legendre equation of degree $\nu=n/p$, which in general is not an integer. Its growing solution is the series
\begin{equation}\label{eq:Fseries}
 F_\nu=X^\nu\sum_{k\geq0}a_kX^{-2k},\qquad a_0=1,\qquad
 a_k=\frac{(\nu-2k+2)(\nu-2k+1)}{2k(2k-2\nu-1)}\,a_{k-1},
\end{equation}
whose coefficients remain regular at integer $\nu$; note $2k-2\nu-1$ is odd. For integer $n\geq0$, $F_n=P_n/[2^{-n}\binom{2n}{n}]$, with $P_n$ the Legendre polynomial normalised by $P_n(1)=1$~\cite{DLMF}. The frozen equation \eqref{eq:frozen} also has the decaying solution $D_n=\mu^{-n/p-1}F_{-n/p-1}(X)\sim r_{\mathrm a}^{-n-p}$. At $p=1$ it is proportional to the Legendre function of the second kind, $F_{-n-1}=\frac{(2n+1)!}{2^n(n!)^2}\,Q_n$, with $Q_0(x)=\atanh(1/x)$. Note that the same construction gives the harmonic radius, $r=B_1(r_{\mathrm a})$, which at $p=1$ reduces to $r=r_{\mathrm a}-\mu$, so that $X=x$.

We now use the frozen solution $B_\ell$ to reconstruct a
coupled scalar--metric perturbation. Substitution of the
following expressions into the linearised equations before
gauge fixing reduces them to the frozen equation for $B_\ell$
and its derivative
(Appendix~\ref{sec:app_inner}):
\[
\begin{gathered}
 v=e=0,\qquad
 c=-\frac{2N\mu}{p}r_{\mathrm a}^{1-p}
       \partial_{r_{\mathrm a}}B_\ell,\qquad
 f=-(p-1)c,\\
 \psi=\lambda B_\ell
 -N\left(1-\frac{\mu}{r_{\mathrm a}^p}\right)
       r_{\mathrm a}\partial_{r_{\mathrm a}}B_\ell .
\end{gathered}
\]
We restore the harmonic condition by adding a pure-gauge
perturbation generated by a vector field $\xi^i$:
\[
 \delta\phi\mapsto
   \delta\phi+\xi^i\partial_i\bar\phi,\qquad
 \delta\gamma_{ij}\mapsto
   \delta\gamma_{ij}+(\mathcal L_\xi\bar\gamma)_{ij}.
\]
The background
fields and the flat reference metric are held fixed in this
transformation of the perturbations. We choose $\xi^i$ so
that the transformed perturbation has
$\delta\mathcal C^i=0$.

The equations for $\xi^i$ reduce to two
inhomogeneous Legendre equations of degrees $n/p$, where
$n=\ell\pm1$ (Appendix~\ref{sec:app_inner}).
Variation of parameters requires the antiderivative
\begin{equation}\label{eq:growingprimitive}
 J_n^+(p,X)=
 \AC\int(X^2-1)F_{n/p}'F_{\ell/p}'F_{1/p}'\,\dd X,
\end{equation}
where primes denote $X$ derivatives. 
Changing the integration constant in $J_n^+$ changes the
coordinate correction $\xi^i$ and hence the reconstructed
perturbation. We fix the antiderivative using $\AC\int$ at
generic $p$, then take $p\to1$. This order can retain a finite
constant that would be missed by setting $p=1$ before integration.

To illustrate how continuation fixes the constant, take
$\ell=2$, $n=1$. As $p\to1$, the integrand tends to
$2X^3-2X$, whose antiderivative is $X^4/2-X^2$
up to a constant.
At generic $p$, the integrand also contains
$[-2(p-1)+O((p-1)^2)]X^{4/p-5}$. Although this term vanishes
at $p=1$, its integral gives
\[
 \frac{-2(p-1)+O((p-1)^2)}{4/p-4}X^{4/p-4}
 \longrightarrow\frac12.
\]
Consequently,
\[
 J_1^+(1,X)=\frac{X^4}{2}-X^2+\frac12
           =\frac{(X^2-1)^2}{2}.
\]
Setting $p=1$ before integration would lose the information
fixing the constant. Without it, the reconstructed column~\eqref{eq:physicalcolumn} below would not belong to the inner family selected at generic dimension (Appendix~\ref{sec:app_inner}).

For general $\ell$ and $n=\ell\pm1$,
Appendix~\ref{sec:app_inner} gives
\begin{equation}\label{eq:endconstant}
 J_n^+(1,X)=\lim_{p\to1}J_n^+(p,X)
 =\int_1^X(t^2-1)F_n'(t)F_\ell'(t)\,\dd t .
\end{equation}
Here $F_{1/p}'\to F_1'=1$, and the lower limit $1$ records
the constant fixed by continuation.

At $p=1$, $X=x=r/\mu$ and $\lambda=\ell(\ell+1)$.
With primes now denoting $x$ derivatives, the reconstructed
column associated with a solution $P$ of $L_\ell P=0$ is
\begin{equation}\label{eq:physicalcolumn}
 q[P]=\begin{pmatrix}
 \Delta P''\\0\\-2\Delta P'/x^2+2\Delta P/x^3\\
 -4P'+2\lambda P/x\\-4P/x
 \end{pmatrix},\quad
 q_{\rm phys}(r,1)=\frac{\mu^\ell}{\ell(\ell-1)}q[F_\ell],
 \quad \text{with} \quad
  \frac{q_{\rm phys}(r,1)}{r^\ell}
\xrightarrow{r\to\infty}
 \begin{pmatrix}1\\0\\0\\0\\0\end{pmatrix}.
\end{equation}
For $q[P]$,
\[
 w_r=w_\Omega=0,\qquad \Psi=\Delta P''.
\]
The solution $q_{\rm phys}$ belongs to the inner family,
used inside the light-source radius.

\subsection{Absence of dimensional poles}\label{sec:poles}
We now show that the coupled radial perturbation $q$ produced
by the light scalar source has no poles at $p=1$, for each
$\ell\geq2$, order by order in the weak-field expansion.

Suppose there is a pole at $p=1$, and let $q_*$ be the
coefficient of the most singular Laurent term. Its pole
order is finite by the uniform bound in
Section~\ref{sec:response}.
Since the operator is analytic and the delta-function source has no pole, the perturbation equation at this Laurent order is homogeneous, so $q_*$ solves the $p=1$ equations and has no source jump. Following the decomposition of Section~\ref{sec:decoupled}, we exclude it in three steps. Its gauge residual vanishes, because the residual of the full response is analytic at $p=1$; its master variable then vanishes, by pairing with $q_{\rm phys}$ and imposing decay at infinity; and the remaining pure-gauge part is excluded by decay, reciprocity, scaling and scalar sign reversal.

Let us start with the gauge residual, i.e.\ the Ward identities (first two equations) in~\eqref{eq:reduced}. At generic $p$, these are frozen equations~\eqref{eq:frozen} of degrees $n=\ell\pm1$, sourced by the force on the light source. The inverse of the frozen operator in~\eqref{eq:frozen} is the Green's function
\begin{equation}
 -\frac{B_n(r_<)\,D_n(r_>)}{2n+p},
\end{equation}
where $r_<$ and $r_>$ are the smaller and larger of $r$ and $R$, as in Section~\ref{sec:prescription}, and $B_n,D_n$ are evaluated at the corresponding areal radii. The coefficients of $B_n$ and $D_n$ are analytic near $p=1$,
since the denominators in~\eqref{eq:Fseries} are nonzero
there for $\nu=n/p$ and $\nu=-n/p-1$. At first order in
$J_2$, the light-source term in the Ward identity is
proportional to $J_2\partial_i\bar\phi(\mathbf x_2)$.
It therefore involves only the known background, whose
weak-field coefficients are regular at $p=1$, and its
harmonic projection introduces no pole, so the frozen inverse applied to this source is analytic
at $p=1$.
A pole could therefore enter the residual only through a homogeneous frozen solution added by the inner or outer prescription, and comparing radial powers excludes this.\footnote{An inner seed of degree $j$ produces residual terms $\mu^kr^{j-1-kp}$, whereas $D_n$ starts as $r^{-n-p}$; at generic $p$ these match only if $j+n-1=(k-1)p$, i.e.\ $k=1$ and $j+n=1$. For $\ell\geq2$, this happens only at $\ell=2$, $j=0$, $n=1$, but there the point-source correction~\eqref{eq:contact} is purely scalar and has no residual. Outside the light source nothing new can arise, since the outer columns contain only decaying powers and the frozen inverse already uses the decaying solution $D_n$ there.} Finally, the Ward identities could also be sourced at the heavy point, by the force on the heavy body. However, a force is a single Cartesian vector, so it contributes only to the dipole and is absent for $\ell\geq2$. The residual of the full response is therefore just the frozen inverse applied to the light-source force, which is analytic at $p=1$. In particular, its coefficient at the most singular Laurent order, the residual of $q_*$, vanishes.

At $p=1$, for a homogeneous solution $q$ with vanishing gauge
residual, the pairing with $q[P]$ reduces to a Wronskian of
master variables:
\begin{equation}\label{eq:physicalpair}
 \B(q[P],q)
 =4\mu\Delta\left[
   (\Delta P'')\Psi'-(\Delta P'')'\Psi
 \right].
\end{equation}
Here primes denote $x$ derivatives. The factor $\mu$ converts
the dimensionless pairing computed with the matrices of
Appendix~\ref{sec:app_rows} to the $r$-normalisation
of~\eqref{eq:pair}.
Inside the light source, both $q_{\rm phys}$ and $q$ belong to the inner family, so their pairing vanishes at every $p$.
We write $q=q_*(p-1)^{-m}+O\bigl((p-1)^{1-m}\bigr)$. Since $q_{\rm phys}$ and the coefficients of the pairing are analytic at $p=1$, the most singular term of $\B(q_{\rm phys},q)$ is $(p-1)^{-m}\,\B\bigl(q_{\rm phys}(1),q_*\bigr)$, evaluated with the $p=1$ coefficients. But as the pairing vanishes for every $p$, this term vanishes, so $\B\bigl(q_{\rm phys}(1),q_*\bigr)=0$.
By~\eqref{eq:physicalpair}, the master variable $\Psi_*$ of $q_*$ is then proportional to the master variable $\Delta F_\ell''$ of $q_{\rm phys}$, which grows. But $q_*$ has no source jump, so the same growing solution continues outside the light source, where decay at infinity forces its coefficient to vanish.

We are left with a solution whose residual and master variable both vanish, i.e.\ a pure-gauge solution preserving the harmonic condition. (Once $v=e=0$ is reached by a coordinate choice, the invariant scalar combination vanishes and the remaining Einstein equations leave no other possibility.) In parallel with the residual in Section~\ref{sec:decoupled}, we describe the coordinate change by the vector field $\xi^i$ that generates it, so that the scalar shifts by $\delta\phi=\xi^i\partial_i\bar\phi$ and the metric by its Lie derivative along $\xi$. We write its components as $\xi^r=\mu\,\xi_rY$ and $\xi^A=\xi_\Omega D^AY$.
It generates the pure-gauge column
\begin{equation}\label{eq:coordinate}
 g[\xi]=\left(\frac{\xi_r}\Delta,\ 2\xi_r',\
 \frac{\xi_r+\Delta\xi_\Omega'}x,\
 \frac{2x\xi_r}\Delta-\lambda\xi_\Omega,\ 2\xi_\Omega\right)^T,
\end{equation}
whose entries are $(\psi,f,v,c,e)$ of the perturbation generated by $\xi$. Thus $q_*=g[\xi]$ for some $\xi$.
Since the displaced Cartesian coordinates must remain harmonic, the displacement obeys the frozen equations~\eqref{eq:frozen} of degrees $\ell\pm1$, and decay of the full metric excludes their growing branches. Note that checking the scalar alone would not suffice at $\ell=2$: there the growing degree-one displacement, linear in $x$, produces a constant metric deformation whose scalar component nevertheless decays. The two surviving solutions have scalar components $Q_{\ell-1}(x)/\Delta$ and $Q_{\ell+1}(x)/\Delta$.

To exclude these, we use three properties of the dimensional Green's function: scaling, reciprocity, and the sign-reversal symmetry of Section~\ref{sec:setup}. Under a positive rescaling $\sigma$,
\begin{equation}\label{eq:scaling}
 \mathcal G(\sigma r,\sigma R;p,\sigma^p\mu)=\sigma^{-p}\mathcal G(r,R;p,\mu),
\end{equation}
so the scalar component $\psi_*(r,R)$ of $q_*$ obeys $(r\partial_r+R\partial_R+\mu\partial_\mu+1)\psi_*=0$; the $p$-dependence of the scaling only affects lower Laurent orders, which may contain logarithms. For fixed $R$, $\psi_*$ is a combination of the two scalar components above. Reciprocity, $\psi_*(r,R)=\psi_*(R,r)$, places its $R$-dependence in the same two-dimensional span. Scaling then fixes the remaining dependence on $\mu$:
\begin{equation}
 \psi_*(r,R)=\frac1\mu\sum_{m,n\in\{\ell-1,\ell+1\}}\kappa_{mn}\,\frac{Q_m(x)\,Q_n(y)}{(x^2-1)(y^2-1)},\qquad \kappa_{mn}=\kappa_{nm}.
\end{equation}
Finally, we use scalar sign reversal. Since $\phi\to-\phi$, $J_a\to-J_a$ leaves the auxiliary theory invariant, $\psi_*$ must be even under $\mu\to-\mu$, understood as an operation on the formal weak-field series, on which $Q_n(-x)=(-1)^{n+1}Q_n(x)$. Both $Q_{\ell\pm1}$ have parity $(-1)^\ell$, so the products above are even, while the factor $1/\mu$ makes the expression odd, and hence $\kappa_{mn}=0$. Since the two pure-gauge solutions have independent scalar components, the whole column $q_*$ vanishes.

Repeating the argument at each remaining negative Laurent order shows that the coupled radial perturbation $q$ produced
by the light scalar source is regular at $p=1$. The same reasoning gives the uniqueness we need in Section~\ref{sec:higher}: two finite candidates with the same gauge residual and master variable differ only by pure-gauge terms, which decay, reciprocity, scaling and sign reversal exclude in exactly the same way. We can therefore construct the higher-multipole Green's function from the simpler $p=1$ equations, provided we impose these conditions.

\section{The finite higher-multipole Green's functions}\label{sec:higher}
Section~\ref{sec:poles} established that, for $\ell\geq2$,
the coupled perturbation produced by the light scalar source
is regular at $p=1$. Once its gauge residual and master
variable are fixed by the dimensional prescription, the
remaining freedom is excluded by requiring the metric to
decay and $\mathcal G_{\psi\psi}$ to be reciprocal, to obey
the scaling~\eqref{eq:scaling} and to be even under
$\mu\to-\mu$. We can therefore construct the response
directly at $p=1$, from the decoupled equations of
Section~\ref{sec:decoupled}, and check these conditions.

\subsection{Legendre Green's functions}
We write the explicit kernels for $x,y>1$, using
$x_<=r_</\mu$ and $x_>=r_>/\mu$. When both radial
arguments occur, we write $\Delta_x=x^2-1$ and
$\Delta_y=y^2-1$.
The decoupled equations~\eqref{eq:reduced} are solved by
\begin{equation}
\label{eq:legendreGF}
 \Rop_n=-P_n(x_<)Q_n(x_>),\qquad
 \Mop_\ell=-\frac{[\Delta P_\ell''](x_<)\,[\Delta Q_\ell''](x_>)}{\lambda(\lambda-2)},
\end{equation}
which invert $L_n$ and $L_\ell-4/\Delta$, respectively. Each uses the growing solution below the source and the decaying one above it, normalised by the conserved Wronskians $\Delta(P_nQ_n'-P_n'Q_n)=-1$ and $\Delta\bigl[(\Delta P_\ell'')(\Delta Q_\ell'')'-(\Delta P_\ell'')'(\Delta Q_\ell'')\bigr]=-\lambda(\lambda-2)$.
Note that $\Rop_n/\mu$ is the $p=1$ limit of the frozen Green's function of Section~\ref{sec:poles}, and that the inner branch of $\Mop_\ell$ is the master variable of $q_{\rm phys}$, since $\Delta F_\ell''\propto\Delta P_\ell''$. Thus $\Rop_{\ell\pm1}$ gives the gauge residual and $\Mop_\ell$ the master variable.

Recovering $\psi$ requires one further step. From the residual and master variable we can build a representative with $v=e=0$, but its own residual need not equal the one fixed by the Ward identities in~\eqref{eq:reduced}. We correct it by adding a pure-gauge column~\eqref{eq:coordinate}, chosen so that the total residual takes the required value. Determining this correction requires a second Legendre inversion, acting on sources that already contain $\Rop_{\ell\pm1}$.
This produces the composed kernels $\mathcal C_n=\Rop_n(3x^2-1)\Rop_n$, i.e.\ two Legendre Green's functions with multiplication by $3x^2-1$ in between. For $x<y$, splitting the intermediate integration at $x$ and $y$ gives
\begin{align}\label{eq:composition}
 \mathcal C_n(x,y)={}&Q_n(x)Q_n(y)\int_0^x(3t^2-1)P_n^2\,\dd t
 +P_n(x)Q_n(y)\int_x^y(3t^2-1)P_nQ_n\,\dd t\notag\\
 &+P_n(x)P_n(y)\int_y^\infty(3t^2-1)Q_n^2\,\dd t,
\end{align}
where unlabelled functions are evaluated at $t$; for $x>y$ we exchange $x$ and $y$. The last integral converges for $n\geq1$, and the first involves only the polynomial $P_n^2$, so $Q_n$ is never needed below $x=1$.

The lower limit of the first integral deserves a comment, since it is not a boundary condition but a choice of integration constant: changing it, for instance to the horizon value $t=1$, adds a multiple of $Q_n(x)Q_n(y)$, which the local equations cannot exclude. Scalar sign reversal (Section~\ref{sec:setup}) fixes the choice. Under $\mu\to-\mu$ we have $x\to-x$, and $K_\ell$ must be odd, so that $\mathcal G_{\psi\psi}=-K_\ell/(4\mu)$ is even. Since the integrand $(3t^2-1)P_n^2$ is even, its primitive vanishing at $t=0$ is odd, making the first term odd, whereas $Q_n(x)Q_n(y)$ is even and therefore excluded.\footnote{The lower limit $1$ excluded here is the one that appears in $J_n^+$, eq.~\eqref{eq:endconstant}. There it was fixed by dimensional continuation, whereas here scalar sign reversal fixes the lower limit $0$; the two prescriptions are not interchangeable.}

On the diagonal, the middle term vanishes, and we define for later convenience
\begin{equation}\label{eq:In}
 I_n(x)\equiv\mathcal C_n(x,x)=Q_n(x)^2\int_0^x(3t^2-1)P_n^2\,\dd t+P_n(x)^2\int_x^\infty(3t^2-1)Q_n^2\,\dd t .
\end{equation}

\subsection{The scalar Green's function}
Combining the master response, the gauge residual of degrees $\ell\pm1$ and the pure-gauge correction, and suppressing the arguments $(x,y)$, we find our desired Green's function for $\ell\geq2$: recalling \eqref{eq:GK}, we have
\begin{align}\label{eq:Khigher}
 K_\ell={}&\Mop_\ell
 -\frac2{2\ell+1}
  \left(\frac1{\Delta_x}+\frac1{\Delta_y}\right)
  (\Rop_{\ell+1}-\Rop_{\ell-1})\notag\\
 &+\frac4{(2\ell+1)\Delta_x\Delta_y}
  (\mathcal C_{\ell+1}-\mathcal C_{\ell-1}),
\end{align}
as derived in Appendix~\ref{sec:app_higher}.
For the unit source $T=\delta(x-y)$, the reconstructed
coupled perturbation has the required source jump and
the gauge residual and master variable constructed above
from $\Rop_{\ell\pm1}$ and $\Mop_\ell$. Its inner master
branch is proportional to that of $q_{\rm phys}$, and its
metric perturbations decay at least as $x^{1-\ell}$ at
infinity.

The scalar function $K_\ell$ is reciprocal and odd under
$\mu\to-\mu$, understood as an operation on the formal
weak-field expansion. Consequently,
$\mathcal G_{\psi\psi}=-K_\ell/(4\mu)$ is reciprocal and
even under mass reversal. Since $K_\ell$ depends only on
$x,y$, the factor $1/\mu$ also gives
$\mathcal G_{\psi\psi}$ the scaling~\eqref{eq:scaling}.
The uniqueness result of Section~\ref{sec:poles} therefore
identifies $-K_\ell/(4\mu)$ with the coefficientwise
$p\to1$ limit of the dimensional Green's function
$\mathcal G_{\psi\psi}(r,R;p)$.
At coincidence, $x=y=1/u$, the contribution of each higher
multipole to the angular sum is
\begin{equation}
 T_\ell(1,u)=-(2\ell+1)\,x\,K_\ell(x,x),
\end{equation}
to be employed in \eqref{eq:CT}.

\section{The two exceptional multipoles}\label{sec:exceptional}
The monopole and dipole cannot be obtained by setting $\ell=0,1$ in Section~\ref{sec:higher}, because the master Green's function contains the factor $1/[\lambda(\lambda-2)]$, and $\lambda(\lambda-2)=0$ for both; recall that $\lambda=\ell(\ell+p)$, with $p=1$ in the physical limit. Instead, we return to the source conditions. The monopole is fixed by requiring that the heavy-source flux $S_r\partial_r\bar\phi=p\mu$ of Section~\ref{sec:background} does not change. The dipole also involves the force on the heavy source, leaving only two constants, which we fix with two exact coefficients computed at generic dimension.

\subsection{Monopole}
For $\ell=0$, only the perturbations $\psi$, $f$ and $c$ of \eqref{eq:fields} are present, and it is convenient to work with the flux change $F$ and the coordinate-invariant scalar $\chi$,
\begin{equation}
 F=\Delta\psi'+c-f/2,\qquad \chi=\psi-c/(2x).
\end{equation}
The scalar equation becomes $F'=T$, so $F$ is constant on either side of the light source. Since the heavy source is held fixed, $F=0$ inside, and the unit jump then gives $F=1$ outside. (Recall from Section~\ref{sec:response} that $S_r\partial_r\bar\phi=p\mu$ is the background flux, whereas $\mu F$ is its linearised perturbation at $p=1$.)

Let us check that the inner solutions selected at generic dimension respect this condition, since they define the Green's function. 
With $r$ derivatives and general $p$, the flux perturbation is $\mu F=S_r\partial_r\psi+p\mu(Nc-f)/2$.
Two inner columns have degree zero and are known exactly: in the order $(\psi,f,c)$, the constant scalar shift and the radial dilation generated by $\xi^r=r/2$,
\begin{equation}
 (1,0,0)^T,\qquad
 \bigl(\tfrac12r\partial_r\bar\phi,\ 1+r\partial_rA/(2A),\ r\partial_rB/(2B)\bigr)^T,
\end{equation}
both of which leave the flux unchanged. The third inner column has degree two and its flux change is constant, but its weak-field powers cannot produce a constant at generic $p$, so it vanishes too. The whole inner family therefore keeps the heavy-source flux fixed.

Taking $p=1$, the gauge residual has only a radial component. As in Section~\ref{sec:poles}, it is the degree-one frozen Green's function applied to the light-source force, and the flux equation then determines $\chi$:
\begin{equation}
 w_r=-\frac{4\Rop_1(x,y)}{\Delta_y},\qquad
 F=x^2\chi'-\frac{x\,w_r}{2}+\frac{\Delta\,w_r'}{4}.
\end{equation}
In these variables, the pole argument is particularly simple. The coefficient of a highest pole has zero flux variation (since neither the fixed heavy flux nor the light-source jump has a pole) and zero residual (since the frozen Green's function is analytic). Hence $\chi'=0$, and decay sets $\chi=0$. What remains is a pure-gauge solution. Decay of the full metric excludes the dilation and leaves the scalar part $Q_1/\Delta$, whose contribution to $\mathcal G_{\psi\psi}$ would, by reciprocity and the scaling~\eqref{eq:scaling}, be proportional to $\mu^{-1}[Q_1/\Delta](x)\,[Q_1/\Delta](y)$. This is odd under $\mu\to-\mu$, so scalar sign reversal excludes it. Repeating the argument at each negative Laurent order, and applying it to the difference of two finite solutions, shows that the monopole is finite and unique under these conditions.

To construct it, we take the representative with $c_{\rm rep}=0$, so that $\psi_{\rm rep}=\chi$, integrate for $\chi$ with $\chi(\infty)=0$, and add the pure-gauge column~\eqref{eq:coordinate}, with $\xi_\Omega=0$, that brings the residual to the value above:
\begin{align}
 f_{\rm rep}&=\frac{\Delta w_r}{x}-\frac{\Delta^2w_r'}{2x^2}-\frac{2F}{x^2},\notag\\
 L_1\xi_r&=-(x^2-x^{-2})w_r+\tfrac12(x^3-x-x^{-1}+x^{-3})w_r'+2(x^{-1}+x^{-3})F,\\
 \psi&=\chi+\frac{\xi_r}{\Delta},\qquad f=f_{\rm rep}+2\xi_r',\qquad c=\frac{2x\xi_r}{\Delta}.\notag
\end{align}
Substitution confirms all three rows of~\eqref{eq:sourced} for $\ell=0$, with the required residual, for any $T$.
With the lower limits of Section~\ref{sec:higher}, the result is reciprocal and its full metric decays. Its coincident value, which enters the angular sum through
\begin{equation}\label{eq:mono}
 T_0(1,u)=-x\,K_0(x,x), \qquad \text{is} \qquad K_0(x,x)=\frac{4I_1}{\Delta^2}+\frac{4x^2Q_0}{\Delta}-\frac{x(5x^2-3)}{\Delta^2},
\end{equation}
with $I_1$, $Q_0$ and $\Delta$ evaluated at $x$.

\subsection{Dipole}\label{sec:dipole}

The dipole differs from the higher multipoles in two ways. First, the Ward identities acquire a second source, the force on the heavy body: its position is held fixed, but the force on it is not zero. Second, once the gauge residual is fixed, symmetry no longer removes all homogeneous solutions: a constant translation gives the pure-gauge scalar $1/\Delta$, which is even under $\mu\to-\mu$, and two constants remain. We determine the force from translation invariance, and the two constants from two weak-field coefficients computed exactly at generic dimension.

Since $W$ is stationary in the fields, a derivative with respect to a source position acts only on the explicit coupling: $\partial W/\partial x_a^i=J_a\partial_i\phi_{\rm reg}(\mathbf x_a)$, where ${\rm reg}$ denotes evaluation at the source with the
same dimensional prescription as $W$. The spatial derivative
is taken before coincidence, and scaleless self-field terms
are set to zero in dimensional regularisation.
Translation invariance, $(\partial_{x_1^i}+\partial_{x_2^i})W=0$, then states that the two forces are equal and opposite. At first order in the light source $\delta J_2$, with $\mathbf x_1=0$ and $\delta\phi$ the scalar response,
\begin{equation}\label{eq:force}
 J_1\partial_i\delta\phi_{\rm reg}(0)=-\delta J_2\,\partial_i\bar\phi(\mathbf x_2).
\end{equation}
This force is a Cartesian vector, so it sources the Ward identities in~\eqref{eq:reduced} only at $\ell=1$, and the gauge residual becomes
\begin{equation}\label{eq:dipward}
 w_-=-\frac4{3\Delta_y}\bigl[\Rop_0(x,y)+Q_0(x)\bigr],
 \qquad w_+=\frac4{3\Delta_y}\Rop_2(x,y),
\end{equation}
where the term $Q_0(x)$ is the heavy-force contribution. Outside the light source, $\Rop_0(x,y)=-Q_0(x)$ since $P_0=1$, so the degree-zero residual vanishes there, as it must for zero total force; at generic $p$ the same cancellation holds because the growing degree-zero frozen solution is a constant. The force also fixes the one inner term that the power counting of Section~\ref{sec:poles} allows for the dipole ($j=1$, $n=0$, $k=1$), so the residual is completely determined, and analytic at $p=1$.

Two solutions with this residual differ by a homogeneous solution with zero residual, and in the dipole such a solution is pure gauge: with $\psi_{\rm rep}=c_{\rm rep}=0$, the metric equations give
\begin{equation}\label{eq:diprep}
 f_{\rm rep}=\Delta\bigl(\Delta w_r'/2+xw_\Omega'-w_\Omega\bigr),\qquad
 xv_{\rm rep}=\tfrac12\bigl[xf_{\rm rep}-\Delta(w_\Omega'+w_r-2w_\Omega/x)\bigr],
\end{equation}
which vanish together with the residual. The pure-gauge solutions whose full metric decays have scalar parts $1/\Delta$, from a constant translation, and $Q_0/\Delta$ and $Q_2/\Delta$. The displacement built from $P_2$ produces a growing metric
perturbation and is excluded. As in Section~\ref{sec:poles}, reciprocity and scaling make any such homogeneous contribution to $\mathcal G_{\psi\psi}=-K_1/(4\mu)$ a symmetric combination of products of these scalar parts, one in each argument. Here, however, scalar sign reversal does not remove everything: $1/\Delta$ is even under $\mu\to-\mu$, while $Q_0/\Delta$ and $Q_2/\Delta$ are odd, so, since $K_1$ must be odd for $\mathcal G_{\psi\psi}$ to be even, only the mixed products of $1/\Delta$ with $Q_0/\Delta$ or $Q_2/\Delta$ survive. Symmetrised, each is proportional to $[Q_n(x)+Q_n(y)]/(\Delta_x\Delta_y)$. Hence, any two such dipole Green's functions with the same source, residual and decay differ by
\begin{equation}\label{eq:dipamb}
 \delta K_1(x,y)=
 \frac{\kappa_0[Q_0(x)+Q_0(y)]+\kappa_2[Q_2(x)+Q_2(y)]}{6\Delta_x\Delta_y},
\end{equation}
with two constants $\kappa_0,\kappa_2$, and a possible highest-pole coefficient has the same form. If $K_{\rm aux}$ is a particular solution, constructed below, then $K_1=K_{\rm aux}+\delta K_1$, and the normalised diagonal $T_1=-3xK_1(x,x)$ at $x=1/u$ (Section~\ref{sec:higher}) is $-3xK_{\rm aux}(x,x)+\delta T_1$, with
\begin{equation}\label{eq:triangular}
 \delta T_1=-\kappa_0u^4-\Bigl(\frac73\kappa_0+\frac2{15}\kappa_2\Bigr)u^6+O(u^8).
\end{equation}
The $u^4$ and $u^6$ coefficients of $T_1$ therefore determine both constants, which is why a finite calculation settles the dipole to all orders.

These two coefficients can be computed exactly at generic dimension. Let us write $T_\ell=\sum_{k\geq0}T_{\ell,k}(p)\,z_R^k$, with $z_R=\mu/R^p$. The recurrence~\eqref{eq:recurrence} and the construction~\eqref{eq:Green} give rational functions $T_{1,4}(p)$ and $T_{1,6}(p)$ (Appendix~\ref{sec:app_dipole}), regular at $p=1$, with
\begin{equation}\label{eq:dipcoeff}
 T_{1,4}(1)=\frac{11901}{6125},\qquad T_{1,6}(1)=\frac{567509}{385875}.
\end{equation}
Their regularity excludes poles: the highest-pole coefficient has the form~\eqref{eq:dipamb}, but since $T_{1,4}$ and $T_{1,6}$ have no pole, it contributes nothing at $u^4$ and $u^6$, so~\eqref{eq:triangular} forces its constants to vanish. The argument repeats through the bounded Laurent orders. The values~\eqref{eq:dipcoeff} then fix $\kappa_0$ and $\kappa_2$ for the finite part.

We now construct a particular solution and compare it with these values. As in Section~\ref{sec:higher}, the light-source part of~\eqref{eq:dipward} gives a Green's function $K_L$ (Appendix~\ref{sec:app_lightdipole}) with diagonal
\begin{equation}\label{eq:lightdip}
 3K_L=\frac{4I_2}{\Delta^2}+\frac{(3x^2-1)^2Q_0}{\Delta}-\frac{x(9x^4-13x^2+2)}{\Delta^2}.
\end{equation}
For the heavy-force contribution, we construct a metric
representative with $\psi_{\rm rep}=0$ and add a coordinate
correction (Appendix~\ref{sec:app_heavyforce}).
After factoring out $-4/(3\Delta_y)$, its radial displacement is
\begin{equation}
 \xi_H=\frac{(7-9x^2)Q_0+9x^3+18x}{36}
\end{equation}
Since the representative has no scalar component, the scalar
contribution is $-4\xi_H(x)/(3\Delta_x\Delta_y)$. Adding the same term with $x$ and $y$ exchanged, which for fixed $y$ is a constant translation and changes neither the residual nor the source jump, gives the reciprocal candidate
\[
 K_{\rm aux}=K_L-\frac{4[\xi_H(x)+\xi_H(y)]}{3\Delta_x\Delta_y}.
\]
Its full metric decays, since the cancellation in~\eqref{eq:dipward} removes the growing terms, so it differs from the dimensionally continued Green's function only by~\eqref{eq:dipamb}. Comparing coefficients,
\begin{center}
\begin{tabular}{lcc}
\toprule
 & $u^4$ & $u^6$\\
\midrule
candidate $-3xK_{\rm aux}(x,x)$ & $229609/55125$ & $7499227/1157625$\\
exact, \eqref{eq:dipcoeff} & $11901/6125$ & $567509/385875$\\
difference & $-20/9=-\kappa_0$ & $-676/135=-\tfrac73\kappa_0-\tfrac2{15}\kappa_2$\\
\bottomrule
\end{tabular}
\end{center}
so that $\kappa_0=20/9$ and $\kappa_2=-4/3$. The coincident value, which enters the angular sum through
\begin{equation}\label{eq:dipfinal}
 T_1(1,u)=-3x\,K_1(x,x), \qquad \text{is} \qquad 3K_1(x,x)=3K_L-\frac{8\xi_H}{\Delta^2}+\frac{(20/9)Q_0-(4/3)Q_2}{\Delta^2},
\end{equation}
with $K_L$, $\xi_H$, $Q_n$ and $\Delta$ evaluated at $x$.

\section{The angular sum}\label{sec:angular}
Although each radial Green's function is finite at $p=1$, the sum \eqref{eq:CT} over multipoles of their coincident values diverges. This is because, at large $\ell$, the normalised contribution $T_\ell$ tends to a nonzero constant, the local self-field. In this section, we subtract this constant from every mode, show that the subtraction leaves the dimensionally regularised answer unchanged, and then show that the subtracted sum may be evaluated directly at $p=1$, without losing a finite contribution from dimensional continuation. As in the radial calculation, all statements hold order by order in the weak-field expansion, that is, in powers of $z_R=\mu/R^p$, which become powers of $u$ at $p=1$.

\subsection{Subtraction of the local term}\label{sec:local}
In terms of the normalised modes~\eqref{eq:Tdef}, the mode sum~\eqref{eq:modesum} reads
\begin{equation}\label{eq:angularweight}
 w_2=\frac{8\pi G_dp}{N\Omega_NR^p}\,\AC\sum_{\ell\geq0}\omega_\ell(p)\,T_\ell(p,z_R),
\end{equation}
where, as in Section~\ref{sec:inner}, $\AC$ denotes meromorphic continuation, here of the sum at fixed weak-field order. At $p=1$, every weight $\omega_\ell$ equals one and $z_R=u$, so~\eqref{eq:angularweight} becomes $w_2=(G/R)\,C(u)$, with $C=[\sum_\ell T_\ell(1,u)]_{\rm DR}$ as in~\eqref{eq:CT}. There, at coincident arguments $x=y=R/\mu=1/u$, Sections~\ref{sec:higher} and~\ref{sec:exceptional} give $T_\ell(1,u)=-(2\ell+1)xK_\ell(x,x)$ for every $\ell$.

The constant that $T_\ell$ approaches at large $\ell$ has a simple origin. At large $\ell$, the radial Green's function is concentrated within a distance of order $R/\ell$ of the light source, so it only probes the coefficients of the scalar equation at $r=R$.

These coefficients are the densitised inverse metric,
\[
 \sqrt{\bar\gamma}\,\bar\gamma^{ij}=A_r\,n^in^j+A_t\,(\delta^{ij}-n^in^j),
\]
where $A_r$ and $A_t$ are its radial and tangential coefficients in Cartesian harmonic coordinates. In terms of the background functions $A,B$ of Section~\ref{sec:response}, $A_r=B^{N/2}/(\sqrt A\,r^{p+1})=S_r/r^{p+1}$ and $A_t=\sqrt A\,B^{N/2-1}/r^{p-1}$. Holding these coefficients fixed at their values at $R$, the coincident value of the radial Green's function at large $\ell$ equals the flat value multiplied by $A_{\rm loc}=[A_r(R)A_t(R)]^{-1/2}$ (Appendix~\ref{sec:app_uniform}), which is therefore the large-$\ell$ limit of $T_\ell$. At $p=1$, $A_r=1-u^2$ and $A_t=1$ at the light body, so $A_{\rm loc}=(1-u^2)^{-1/2}$. Mixing with the metric changes $T_\ell$ only at order $\ell^{-2}$, as Lemma~\ref{lem:tail} in the next subsection shows.

As in Section~\ref{sec:exceptional}, we expand $T_\ell=\sum_kT_{\ell,k}z_R^k$, and similarly $A_{\rm loc}=\sum_kA_{{\rm loc},k}z_R^k$. At generic $p$, the coefficients $A_{{\rm loc},k}$ follow from the background recurrence in Appendix~\ref{sec:app_radial}, or equivalently from the areal form of Section~\ref{sec:inner}.

Subtracting $A_{\rm loc}$ from every mode $T_\ell$ might seem to change the answer, but it does not, because the continued sum of the weights vanishes. With an auxiliary convergence parameter $0\leq z_{\rm A}<1$, their generating function is
\begin{equation}\label{eq:weightsum}
 \sum_{\ell\geq0}\omega_\ell(p)z_{\rm A}^\ell=\frac{(1-z_{\rm A})^{-p}}p .
\end{equation}
For $-1<\Re p<0$ the series converges absolutely at $z_{\rm A}=1$, where the right-hand side vanishes, so by meromorphic continuation $\AC\sum_\ell\omega_\ell(p)=0$ for every $p$. Since $A_{\rm loc}$ does not depend on $\ell$, subtracting it changes the continued sum in~\eqref{eq:angularweight} by $A_{\rm loc}\,\AC\sum_\ell\omega_\ell=0$. (At $p=1$, this gives $[\sum_{\ell\geq0}1]_{\rm DR}=0$.)

\subsection{The dimensional limit and EFT matching}
Two points remain before the subtracted sum can be evaluated at $p=1$: we must justify taking the limit $p\to1$ inside the sum, and we must check that the continued sum agrees with the EFT definition of $w_2$, i.e.\ the coincident value of the scalar Green's function computed directly as a $d$-dimensional integral (the Born series of Section~\ref{sec:response}), before any decomposition into multipoles.

The first point requires some care. Suppose the subtracted modes contained a remainder proportional to $(p-1)/\ell$. Each term would vanish at $p=1$, but with the weights~\eqref{eq:omegal}, $\omega_\ell\sim\ell^{p-1}/\Gamma(p+1)$, the continued sum of $\ell^{p-2}$ is the Riemann zeta function $\zeta(2-p)$, which has a pole there. Since $(p-1)\zeta(2-p)\to-1$, this would leave a finite contribution that is invisible mode by mode. We exclude this by showing that the remainder has no $1/\ell$ term even away from $p=1$.

\begin{lemma}\label{lem:tail}
For each fixed weak-field order $k$ there is a neighbourhood $U$ of $p=1$ in the complex plane and a constant $C_k$ such that
\begin{equation}\label{eq:tail}
 \bigl|T_{\ell,k}(p)-A_{{\rm loc},k}(p)\bigr|\leq\frac{C_k}{\ell^2}\qquad\text{for all }\ell\geq1,\ p\in U,
\end{equation}
and the difference $T_{\ell,k}(p)-A_{{\rm loc},k}(p)$ is analytic in $p$ on $U$.
\end{lemma}

The proof is given in Appendix~\ref{sec:app_uniform}. In brief, the leading large-$\ell$ behaviour is that of the scalar operator with its coefficients held fixed at the source, which gives $A_{\rm loc}$; coupling to the metric contributes only at order $1/\ell^2$. In the local large-$\ell$ expansion, the only possible $1/\ell$ correction, from the first variation of the scalar coefficients about the source, is odd under reflection about the source and integrates to zero. Multiplied by the weights, the subtracted terms are therefore $O(\ell^{\Re p-3})$, summable uniformly near $p=1$, so the limit $p\to1$ may be taken term by term.

Let us now turn to the second point, the identification of the continued sum with the EFT integral. We first show that the two agree where both converge, and then continue the equality to $p=1$. In the strip $-1<\Re p<0$, the coincident angular sum converges absolutely, since $\omega_\ell T_\ell=O(\ell^{\Re p-1})$, so the angular expansion of the field can be summed at the source. The angular normalisation is that of dimensional Fourier integration, and the radial integrals and contact terms were already matched in Section~\ref{sec:response}. Regulating the remaining divergences, at the heavy point and at infinity, separately during this comparison, the field integral and the modal sum agree on an open set in this strip. We then continue this equality to $p=1$ along a path in complex $p$ that avoids the poles of the fixed-order coefficients and along which the large-$\ell$ estimate continues to hold (Appendix~\ref{sec:app_continuation}); the finitely many remaining modes are continued individually, their limits being regular by Sections~\ref{sec:continuation} and~\ref{sec:exceptional}. Uniqueness of meromorphic continuation then identifies the modal sum with the EFT integral. At the physical dimension we obtain, coefficient by coefficient in the $u$ expansion,
\begin{equation}\label{eq:sumrule}
 C(u)=\sum_{\ell=0}^\infty\left[T_\ell(1,u)-(1-u^2)^{-1/2}\right].
\end{equation}
In Section~\ref{sec:summation}, we will also use the continued sum over the higher modes alone, $[\sum_{\ell\geq2}T_\ell]_{\rm DR}$, defined as the limit $p\to1$ of $\AC\sum_{\ell\geq2}\omega_\ell(p)T_\ell(p,z_R)$. Subtracting and adding $A_{\rm loc}$ in each mode, and using~\eqref{eq:weightsum} with $\omega_0=1/p$ and $\omega_1=1$, we find
\[
 \Bigl[\sum_{\ell\geq2}T_\ell\Bigr]_{\rm DR}=\sum_{\ell\geq2}\bigl[T_\ell(1,u)-A_{\rm loc}\bigr]-2A_{\rm loc},
\]
so that, by~\eqref{eq:sumrule},
\[
 C(u)=T_0(1,u)+T_1(1,u)+\Bigl[\sum_{\ell\geq2}T_\ell(1,u)\Bigr]_{\rm DR}.
\]
So the low modes may be added without subtraction when the higher modes are summed by continuation.

\section{Closed-form response and potential coefficients}\label{sec:summation}
We finally sum the modes and obtain the first-self-force static potential. The radial expressions are lengthy, but most of their terms cancel in the angular sum. In particular, we will not need to evaluate the composed integrals $I_n$ of~\eqref{eq:In} individually. We first compute the leading weak-field coefficient, which already displays the mechanism, and then perform the sum to all orders. At coincidence, $x=y=1/u$ and $T_\ell=-(2\ell+1)xK_\ell(x,x)$, so that
\[
 C=-x\Bigl[\sum_{\ell\geq0}(2\ell+1)K_\ell(x,x)\Bigr]_{\rm DR},
\]
with the prescription of Section~\ref{sec:angular}. Unless indicated otherwise, radial arguments below are coincident, $(x,x)$.

\subsection{Leading weak-field coefficient}\label{sec:leading}
At $p=1$, $z_R=u$ and $A_{\rm loc}=(1-u^2)^{-1/2}=1+u^2/2+O(u^4)$. In the subtracted sum~\eqref{eq:sumrule}, the flat terms $T_{\ell,0}=1$ cancel against $A_{{\rm loc},0}=1$ mode by mode, and scalar sign reversal makes $C$ even in $u$. Writing $[u^2]C$ for the coefficient of $u^2$, the first nonzero coefficient is therefore
\[
 [u^2]C=\sum_{\ell\geq0}\Bigl(T_{\ell,2}(1)-\frac12\Bigr).
\]
The exceptional modes follow from~\eqref{eq:mono} and~\eqref{eq:dipfinal}:
\[
 T_{0,2}(1)=\frac1{15},\qquad T_{1,2}(1)=\frac{69}{35}.
\]
For the higher modes, we expand the ingredients of~\eqref{eq:Khigher} at large $x$. For $n\geq1$, the Legendre and composed Green's functions give
\[
 \Rop_n(x,x)=-\frac1{(2n+1)x}+O(x^{-3}),\qquad
 I_n(x)=\frac{6x}{(2n-1)(2n+1)(2n+3)}+O(x^{-1}),
\]
while expanding $\Delta P_\ell''$ and $\Delta Q_\ell''$ gives, for the master Green's function,
\[
 -(2\ell+1)x\Mop_\ell(x,x)=1+\frac{2\ell^2+2\ell-9}{(2\ell-1)(2\ell+3)}\,x^{-2}+O(x^{-4}).
\]
Inserting these into~\eqref{eq:Khigher}, we find
\begin{equation}\label{eq:leading_modes}
 T_{\ell,2}(1)-\frac12
 =3\left(\frac1{2\ell-3}-\frac1{2\ell+5}\right)
 -\frac{31}{8}\left(\frac1{2\ell-1}-\frac1{2\ell+3}\right),
 \qquad \ell\geq2.
\end{equation}
Written this way, both differences beautifully telescope, leaving only their lowest-degree terms:
\[
 \sum_{\ell\geq2}\Bigl(T_{\ell,2}(1)-\frac12\Bigr)
 =3\Bigl(1+\frac13+\frac15+\frac17\Bigr)-\frac{31}{8}\Bigl(\frac13+\frac15\Bigr)=\frac{311}{105}.
\]
Adding the exceptional modes gives
\begin{equation}\label{eq:leading_sum}
 [u^2]C=\Bigl(\frac1{15}-\frac12\Bigr)+\Bigl(\frac{69}{35}-\frac12\Bigr)+\frac{311}{105}=4,
\end{equation}
so that $\overline\Gamma_{2,2}=2!\,[u^2]C=8$, the known 2PN value. Notice the role of the common subtraction: each exceptional mode contributes its own $-1/2$, just as the higher modes do. The telescoping between neighbouring degrees persists to all orders, as we now show.

\subsection{Evaluation of the angular sum}
The angular weight $2\ell+1$ cancels the denominator in the ${\mathcal R}$ and ${\mathcal C}$ terms of~\eqref{eq:Khigher}, leaving differences between degrees $\ell+1$ and $\ell-1$. For an integer cutoff $L\geq2$, for example,
\[
 \sum_{\ell=2}^{L}(\Rop_{\ell+1}-\Rop_{\ell-1})
 =\Rop_L+\Rop_{L+1}-\Rop_1-\Rop_2.
\]
The same identity holds with $I_n$ in place of $\Rop_n$. The upper terms vanish coefficientwise as $L\to\infty$, since $\Rop_n=O(n^{-1})$ and $I_n=O(n^{-3})$. The higher-mode sum reduces to
\begin{equation}\label{eq:highsum}
 \left[\sum_{\ell\geq2}(2\ell+1)K_\ell\right]_{\rm DR}
 =\left[\sum_{\ell\geq2}(2\ell+1)\Mop_\ell\right]_{\rm DR}
 +\frac{4(\Rop_1+\Rop_2)}{\Delta}
 -\frac{4(I_1+I_2)}{\Delta^2}.
\end{equation}

Only the master Green's functions still require an infinite sum. To evaluate it, we temporarily separate the field point, at rescaled radius $x$, from the light source, at rescaled radius $y$, also in direction: if their directions make an angle $\theta$, with $\zeta=\cos\theta$, the addition theorem weights each mode by $P_\ell(\zeta)$, and the resulting sum over $\ell$ converges for $-1<\zeta<1$. We seek a generating function whose degree-$\ell$ Legendre coefficient is $-(2\ell+1)\Mop_\ell(x,y)$ for $\ell\geq2$, and then let $\zeta\to1$ and $y\to x$. Since the master operator $L_\ell-4/\Delta$ of~\eqref{eq:reduced} is the associated Legendre operator of order two, we construct it as an angular integral weighted by $\cos(2\varphi)$, with $\varphi$ an auxiliary integration angle, and then verify its properties. We define
\begin{equation}
 \alpha_0=xy-\zeta,\qquad \beta_0^2=(x^2-1)(y^2-1),\qquad
 D=\sqrt{\alpha_0^2-\beta_0^2}.
\end{equation}
Here, $D$ vanishes only when the two points coincide. For $\zeta=1$, it reduces to $|x-y|$, and $1/D$ will carry the local singularity of the sum. For $x,y>1$ away from coincidence, take the positive square root for $\beta_0$ and define the generating function $\mathcal A_2$ by the elementary integral
\[
 \mathcal A_2(x,y,\zeta)=\frac1\pi\int_0^\pi
       \frac{\cos(2\varphi)}{\alpha_0-\beta_0\cos\varphi}\,\dd\varphi
 =\frac{(\alpha_0-D)^2}{\beta_0^2D}.
\]

Away from coincidence, direct differentiation shows that
\[
 \Bigl[\partial_x\Delta_x\partial_x-\frac4{\Delta_x}+\partial_\zeta(1-\zeta^2)\partial_\zeta\Bigr]\mathcal A_2=0 .
\]
Writing $\mathcal A_2=\sum_\ell a_\ell(x,y)P_\ell(\zeta)$ and using $\partial_\zeta(1-\zeta^2)\partial_\zeta P_\ell=-\ell(\ell+1)P_\ell$, each coefficient obeys $(L_\ell-4/\Delta_x)\,a_\ell=0$, the master equation of~\eqref{eq:reduced}, and likewise in $y$, since $\mathcal A_2$ is symmetric in $x$ and $y$.

For $\ell\geq2$, the behaviour of $\mathcal A_2$ for $x<y$ and for $x>y$ selects the branches $\Delta P_\ell''$ and $\Delta Q_\ell''$ fixed in Sections~\ref{sec:continuation}--\ref{sec:higher}, so that $a_\ell\propto\Mop_\ell$.

The normalisation comes from the coincidence: the local singularity $1/D$ makes the left-hand side above equal to $-2\delta(x-y)\delta(\zeta-1)$, and since $\delta(\zeta-1)=\sum_\ell\frac{2\ell+1}2P_\ell(\zeta)$, this gives $(L_\ell-4/\Delta_x)\,a_\ell=-(2\ell+1)\,\delta(x-y)$. Hence
\begin{equation}\label{eq:Acoeff}
 a_\ell(x,y)=-(2\ell+1)\Mop_\ell(x,y),\qquad \ell\geq2.
\end{equation}

The generating function also contains degree-zero and degree-one solutions of the master equation in~\eqref{eq:reduced}. These are not the physical monopole and dipole of the coupled system, which were calculated separately in Section~\ref{sec:exceptional}, so we subtract them. Their coefficients factorise, as in~\eqref{eq:legendreGF}, into an inner solution, evaluated at the smaller of $x$ and $y$, and an outer solution, evaluated at the larger; these are, respectively,
\[
 \ell=0:\ \ \frac{x-1}{2(x+1)}\ \text{and}\ \frac{2x}{\Delta},\qquad
 \ell=1:\ \ \frac{(x-1)(x+2)}{6(x+1)}\ \text{and}\ \frac{2}{\Delta}.
\]
At coincidence, including the multiplicity $2\ell+1$, their sum is
\begin{equation}\label{eq:auxiliary_low_sum}
 \underbrace{\frac{x-1}{2(x+1)}\frac{2x}{\Delta}}_{\ell=0}
 +\underbrace{3\frac{(x-1)(x+2)}{6(x+1)}\frac2\Delta}_{\ell=1}
 =\frac2{x+1}.
\end{equation}

Finally, we take the coincidence limit, $y=x$ and $\zeta\to1$, in the dimensional prescription of Section~\ref{sec:angular}. Setting $y=x$ and $\zeta=1-\eta$, the generating function behaves as
\[
 \mathcal A_2=\frac1D-\frac2\Delta+O(\sqrt\eta),\qquad D=\sqrt{2\Delta\eta+\eta^2}.
\]
Since $T_\ell=-(2\ell+1)xK_\ell$, subtracting $A_{\rm loc}$ from each $T_\ell$ amounts to subtracting $A_{\rm loc}/x=1/\sqrt\Delta$ from each coefficient of $\mathcal A_2$. Summed with the angular weights, using the Legendre generating function $\sum_{\ell\geq0}P_\ell(\zeta)\,t^\ell=(1-2\zeta t+t^2)^{-1/2}$ in the limit $t\to1$, this subtracted term is $(1/\sqrt\Delta)\sum_\ell P_\ell(\zeta)=1/\sqrt{2\Delta\eta}$, so that
\[
 \mathcal A_2-\frac1{\sqrt{2\Delta\eta}}
 =\left(\frac1D-\frac1{\sqrt{2\Delta\eta}}\right)-\frac2\Delta+O(\sqrt\eta)
 \longrightarrow-\frac2\Delta .
\]
Removing $1/D$ therefore agrees with the prescribed subtraction in the coincidence limit, although not at finite angular separation. The large-$\ell$ estimate of Lemma~\ref{lem:tail} applies equally to the master operator, whose additional derivative-free term $-4/\Delta$ changes the normalised response only at order $\ell^{-2}$. The subtracted series therefore converges absolutely, coefficient by coefficient in $u$, so the limit may be taken term by term: $-2/\Delta$ is the dimensionally regularised sum of the coefficients of $\mathcal A_2$ over all $\ell\geq0$. As at the end of Section~\ref{sec:angular}, the low modes can then be removed without subtraction. Removing the auxiliary ones~\eqref{eq:auxiliary_low_sum}, and using~\eqref{eq:Acoeff} for $\ell\geq2$, we obtain
\begin{equation}\label{eq:mastersum}
 \left[\sum_{\ell\geq2}(2\ell+1)\Mop_\ell\right]_{\rm DR}=\frac2\Delta+\frac2{x+1}=\frac{2x}{\Delta}.
\end{equation}

We can now combine this result with the monopole~\eqref{eq:mono}, the weighted dipole~\eqref{eq:dipfinal} and the remainder of~\eqref{eq:highsum}. Remarkably, the composed integrals cancel without ever being evaluated, although they were needed to construct the individual radial responses. Their coefficients are
\begin{center}
\begin{tabular}{lcc}
\toprule
contribution & coefficient of $I_1$ & coefficient of $I_2$\\
\midrule
monopole $K_0$, \eqref{eq:mono} & $4/\Delta^2$ & $0$\\
light dipole $3K_L$, \eqref{eq:lightdip} & $0$ & $4/\Delta^2$\\
higher-mode sum, \eqref{eq:highsum} & $-4/\Delta^2$ & $-4/\Delta^2$\\
\bottomrule
\end{tabular}
\end{center}
The elementary parts simplify just as much. Recall that $K_L$ is the light-source part of the dipole Green's function, with diagonal~\eqref{eq:lightdip}, so that $3K_1$ is $3K_L$ plus the heavy-force and constant terms of~\eqref{eq:dipfinal}. With $\Rop_1(x,x)=-x(xQ_0-1)$ and $\Rop_2(x,x)=-(3x^2-1)Q_2/2$, we find
\[
 \left(K_0-\frac{4I_1}{\Delta^2}\right)
 +\left(3K_L-\frac{4I_2}{\Delta^2}\right)
 +\frac{4(\Rop_1+\Rop_2)}\Delta=0,
\]
i.e.\ the monopole and the light-source part of the dipole cancel the lowest-degree terms left by telescoping. Only the master sum~\eqref{eq:mastersum}, the heavy-force term and the two dipole constants of~\eqref{eq:dipfinal} remain:
\begin{equation}
 C=-x\left[\frac{2x}{\Delta}
 +\frac{-8\xi_H+(20/9)Q_0-(4/3)Q_2}{\Delta^2}\right]
 =\frac{x\,[4x-(4/3)\atanh(1/x)]}{(x^2-1)^2}.
\end{equation}
In the last equality, we used $Q_2=(3x^2-1)Q_0/2-3x/2$. Finally, $u=1/x$ gives
\begin{equation}\label{eq:Cresult}
 {C(u)=\frac{4u^2}{(1-u^2)^2}
                  \left(1-\frac u3\atanh u\right).}
\end{equation}

\subsection{Connected correlators and the static potential}
We can now read off all the connected correlators $\overline\Gamma_{2,2n}$ from this expression. By~\eqref{eq:Cdef}, $[u^{2n}]C=\overline\Gamma_{2,2n}/(2n)!$ for positive integer $n$. Expanding $(1-u^2)^{-2}=\sum_{k\geq0}(k+1)u^{2k}$ and $u\atanh u=\sum_{j\geq1}u^{2j}/(2j-1)$, with integer summation indices $j,k$, gives
\begin{equation}
 [u^{2n}]C=4n-\frac43\sum_{j=1}^{n-1}\frac{n-j}{2j-1},
\end{equation}
which, on using $j/(2j-1)=1/2+1/[2(2j-1)]$, becomes
\begin{equation}\label{eq:gamma}
 {\frac{\overline\Gamma_{2,2n}}{(2n)!}
 =\frac{14n-2}{3}-\frac{4n-2}{3}
       \sum_{j=1}^{n-1}\frac1{2j-1},\qquad n\geq1.}
\end{equation}
Here, an empty sum is zero. For instance, at $n=4$ the sum is $1+\tfrac13+\tfrac15=\tfrac{23}{15}$, giving $[u^8]C=18-\tfrac{14}3\cdot\tfrac{23}{15}=\tfrac{488}{45}$ and $\overline\Gamma_{2,8}=8!\cdot\tfrac{488}{45}=437248$.

Substituting~\eqref{eq:Cresult} into the source-to-mass conversion~\eqref{eq:v1fromC} gives the potential~\eqref{eq:v1SF} and its expansion~\eqref{eq:v1series}. To make the conversion~\eqref{eq:v1fromC} explicit at low order, recall that $C=4u^2+O(u^4)$ enters the potential multiplied by $u$, whereas the heavy-source shift contributes already at $u^2$. The two terms in~\eqref{eq:v1fromC} give
\[
 v_{1SF}(u)=
 \underbrace{\left(\frac{u^2}{2}-u^3+O(u^4)\right)}_{\text{heavy-source shift}}
 +\underbrace{\left(-2u^3+O(u^4)\right)}_{\text{response }C}
 =\frac{u^2}{2}-3u^3+O(u^4).
\]
We see that the new field calculation first contributes at $u^3$, and the leading $u^2/2$ comes entirely from the heavy-source shift.
 For PN order $k$, denote the quadratic-light-mass term by $V_{k\mathrm{PN}}^{(m^2)}$. The first two coefficients in this mass sector beyond ref.~\cite{Brunello} are
\begin{equation}\label{eq:PNpredictions}
 V_{8\mathrm{PN}}^{(m^2)}=-\frac{1574}{45}\frac{G^9M^8m^2}{R^9},
 \qquad
 V_{9\mathrm{PN}}^{(m^2)}=\frac{429}{10}\frac{G^{10}M^9m^2}{R^{10}}.
\end{equation}
They also follow by substituting $\overline\Gamma_{2,8}=437248$ into Eqs.~(B1)--(B2) of ref.~\cite{Brunello}, providing a check on the source combinatorics.

\section{Discussion}\label{sec:discussion}
We have obtained the static two-body potential at first self-force order in closed form, eq.~\eqref{eq:v1SF}, and with it the connected correlators $\overline\Gamma_{2,2n}$ to all orders, eq.~\eqref{eq:gamma}.  The calculation reduces to the coincident response of a system of coupled perturbations on the Schwarzschild background in harmonic gauge. Two features made it tractable. First, the dimensionally regulated point-source theory fixes the relevant Green's functions completely, provided that the limit $d\to3$ is taken after the radial integrations. The finite constant in~\eqref{eq:endconstant} shows that this order matters. Second, in the sum over angular multipoles, terms from neighbouring multipoles telescope and the most complicated parts (composed radial integrals) cancel between the monopole and dipole on one side and the higher multipoles on the other, leaving a single final sum that can be evaluated in closed form.

The result holds coefficient by coefficient in the weak-field expansion, and it concerns the static sector in harmonic gauge. The closed form also shows that the self-force expansion of the static potential breaks down near the horizon, where $(m/M)v_{1SF}$ becomes comparable to $v_{0SF}$.

A natural next question is whether the same strategy extends beyond first self-force order. At second self-force order, the new input consists of the connected correlators with three light-source insertions, or equivalently the response at second order in the light source, which is a more complicated object than the Green's function constructed here. It would also be interesting to extend the method to the velocity-dependent part of the potential at first self-force order, where the response is no longer static. Another direction, which we expect to report on soon, is a spinning heavy body. In each case, a key step will again be the one encountered here: identifying, at generic dimension, which homogeneous solutions the point-source theory selects, thereby fixing the Green's functions.

\section*{Acknowledgements}
\addcontentsline{toc}{section}{Acknowledgements}
I am very grateful to Riccardo Gonzo for collaboration in the early stages of this project. I am supported by a Royal Society University Research Fellowship. Large language models were used substantially in this work: the proof was developed with ChatGPT (OpenAI), and the calculations were checked independently with Claude (Anthropic). Both were also used to assist the writing. I also thank OpenAI and Anthropic for providing academic access to ChatGPT and Claude, respectively. I take full responsibility for the content of this paper.

\appendix
\section{Radial operators and solution reconstruction}\label{sec:app_radial}
This appendix collects the operator data needed for the recurrence, the dimensional construction of the inner solution~\eqref{eq:physicalcolumn}, and the reconstruction algebra for the higher multipoles.

\subsection{Dimensional recurrence}\label{sec:app_recurrence}
We work at generic $p$, with $N=p+1$ and $\lambda=\ell(\ell+p)$, and order the fields as $(\psi,f,v,c,e)$. With $\alpha$ the trial exponent and the abbreviations $t=\alpha(\alpha+p)-\lambda$ and $\eta=p(\lambda-N)/N$, the flat matrix $\mathsf H_0(\alpha,p)$ of~\eqref{eq:powerH} is
\begin{equation}
 \mathsf H_0=\begin{pmatrix}
 -2Nt/p&0&0&0&0\\
 0&-(t-4N)/4&-2\lambda&N(t-4)/4&0\\
 0&-2\lambda&-\lambda(t-4)&2\lambda&-2\lambda\eta\\
 0&N(t-4)/4&2\lambda&N[(N-2)t+4]/4&0\\
 0&0&-2\lambda\eta&0&-\lambda\eta(t+2p)/2
 \end{pmatrix}.
\end{equation}
For $\ell\geq2$, with $\mathsf K_0=-\partial_\alpha^2\mathsf H_0/2$, the angular matrix $\mathsf K_0^{-1}\mathsf H_0(0)$ satisfies the cubic polynomial with roots $(\ell-2)(\ell+p-2)$, $\lambda$ and $(\ell+2)(\ell+p+2)$, which are $j(j+p)$ for the flat exponents~\eqref{eq:flatpowers}; for $\ell=0,1$ the absent components are removed first. The first-order background correction in~\eqref{eq:powerH} is
\begin{equation}
 \mathsf H_1=\begin{pmatrix}
 0&N\alpha&-2N\lambda&-N^2\alpha&0\\
 -N\alpha&0&0&0&0\\
 -2N\lambda&0&0&0&0\\
 N^2\alpha&0&0&0&0\\
 0&0&0&0&0
 \end{pmatrix}.
\end{equation}
Higher matrices are obtained by expanding the radial Hessian derived from the bulk action~\eqref{eq:Sbulk}. To do so, we first expand the background in $z=\mu/r^p$, in terms of the radial and tangential coefficients $A_r,A_t$ of the densitised inverse metric $\sqrt{\bar\gamma}\,\bar\gamma^{ij}$ (Section~\ref{sec:local}). The background field equations and the harmonic condition give
\begin{equation}
 A_t=A_r-\frac pN z\partial_zA_r,\qquad
 \bigl[-pzA_r\partial_zA_t+NA_t^2+(p-1)A_tA_r\bigr]^2=4p^2A_t^2\bigl(z^2+A_tA_r\bigr).
\end{equation}
We write $A_r=1+\sum_{n\geq2,\ n\ {\rm even}}A_{r,n}z^n$. At even order $n$, let $R_n$ be the coefficient of $z^n$ in the left-hand side minus the right-hand side of the second equation, with $A_r$ truncated before order $n$. Then
\begin{equation}\label{eq:Arn}
 A_{r,n}=-\frac{NR_n}{4np^2[(n-1)p-2]}.
\end{equation}
At $p=1$, the recursion stops at $A_r=1-z^2$, with $A_t=1$. The background functions $A,B$ of Section~\ref{sec:background} and $\partial_r\bar\phi$ follow from $A=(A_rA_t)^{1/p}A_t/A_r$, $B=r^2(A_rA_t)^{1/p}$ and $\partial_r\bar\phi=pz/(rA_r)$; at $p=1$ these reproduce~\eqref{eq:background}. The matrices $\mathsf H_k$ then follow by applying the Hessian to $r^\alpha\boldsymbol v$: after factoring out $r^\alpha$, $r\partial_r$ acts as $\alpha-pz\partial_z$ on functions of $z$. These steps generate every matrix from the bulk action by finite rational operations; the ancillary code implements them through sixth order in $\mu$.

\subsection{Physical-dimensional operator}\label{sec:app_rows}
At $p=1$, with $\lambda=\ell(\ell+1)$, $x$, $\Delta$ and $L_\ell$ as in~\eqref{eq:xy}, and primes denoting $x$ derivatives, we write the field equations as $E_\psi=T$ and $E_f=E_v=E_c=E_e=0$; the first is the scalar equation of Section~\ref{sec:decoupled}. The left-hand sides are
\begin{align}
 E_f&=L_\ell f+4\lambda x^2v/\Delta-4(x^2+1)(f-c)/\Delta+8\psi',\notag\\
 E_v&=\Delta(xv)''-(\lambda+4-6/x^2)xv
 +(2x-1/x)f-2xc+(\lambda-2)\Delta e/x+4\psi,\notag\\
 E_c&=L_\ell c-2\lambda v+2(x^2+1)(f-c)/\Delta,\notag\\
 E_e&=L_\ell e+2e+4v,\qquad E_\psi=L_\ell\psi+c'-f'/2+\lambda xv/\Delta.
\end{align}
In terms of these left-hand sides, the Hessian of~\eqref{eq:sourced} is
\begin{equation}
 Hq=\left(-4E_\psi,\,-E_f/4+E_c/2,\,-\lambda xE_v/\Delta,\,
                  E_f/2,\,-\lambda(\lambda-2)E_e/4\right)^T.
\end{equation}
In the form~\eqref{eq:Hform}, the pairing uses
\begin{equation}
 \mathsf K=\begin{pmatrix}
 4\Delta&0&0&0&0\\
 0&\Delta/4&0&-\Delta/2&0\\
 0&0&\lambda x^2&0&0\\
 0&-\Delta/2&0&0&0\\
 0&0&0&0&\lambda(\lambda-2)\Delta/4
 \end{pmatrix},
 \qquad \mathsf S_{\psi f}=2,\quad \mathsf S_{\psi c}=-4,
\end{equation}
with opposite skew entries and all other independent entries zero. These matrices give the pairing in $x$; the $r$-normalised pairing~\eqref{eq:pair} is $\mu$ times this expression.

\subsection{Dimensional reconstruction of the inner solution}\label{sec:app_inner}
This subsection supplies the construction used in Section~\ref{sec:inner}. We return to generic $p$, with $N=p+1$ and $\lambda=\ell(\ell+p)$. We first work in areal radius $r_{\mathrm a}$, with the background~\eqref{eq:arealbackground}, the frozen solutions $B_n(r_{\mathrm a})$ of~\eqref{eq:frozen} and the functions $F_\nu$ of~\eqref{eq:Fseries}. We use the decomposition~\eqref{eq:fields} in $r_{\mathrm a}$, then transform to the harmonic radial coordinate.

The field equations before gauge fixing admit a representative built from the frozen scalar solution, which does not yet have zero gauge residual. It has $v=e=0$ and, with primes denoting $r_{\mathrm a}$ derivatives,
\begin{equation}\label{eq:physicalrep}
 c=-\frac{2N\mu}{p}r_{\mathrm a}^{1-p}B_\ell',\qquad f=-(p-1)c,\qquad
 \psi=\lambda B_\ell-N\left(1-\frac{\mu}{r_{\mathrm a}^p}\right)r_{\mathrm a}B_\ell'.
\end{equation}
The Einstein equations reduce to
\begin{equation}
 r_{\mathrm a}^2e^{2\bar\phi}c''+r_{\mathrm a}\bigl[2p+(p+1)e^{2\bar\phi}\bigr]c'-(\lambda-2p^2)c=0,
\end{equation}
which follows by differentiating the frozen equation~\eqref{eq:frozen}. The angular trace yields this relation without division by $p-1$, so the construction remains valid at the physical dimension.

Restoring zero gauge residual requires a coordinate change, which we write as
\begin{equation}\label{eq:restoring_dictionary}
 \xi^{r_{\mathrm a}}=\frac{\ell h_{\ell-1}+(\ell+p)h_{\ell+1}}{(2\ell+p)B_1'}\,Y,\qquad
 \xi^A=\frac{h_{\ell-1}-h_{\ell+1}}{(2\ell+p)B_1}\,D^AY,
\end{equation}
with $B_1$ and the $h_n$ evaluated at $r_{\mathrm a}$. It adds $\xi^{r_{\mathrm a}}\bar\phi'$ to the scalar perturbation and the Lie derivative $\mathcal L_\xi\bar\gamma$ to the metric perturbation. After transforming to $r=B_1(r_{\mathrm a})$, the resulting column, divided by $\ell(\ell-1)$, has the normalisation of~\eqref{eq:physicalcolumn}. The radial functions $h_n$, of degrees $n=\ell\pm1$, satisfy
\begin{equation}
 \partial_{r_{\mathrm a}}(r_{\mathrm a}^{p+1}e^{2\bar\phi}h_n')-n(n+p)r_{\mathrm a}^{p-1}h_n
 =-2N\mu\partial_{r_{\mathrm a}}(r_{\mathrm a}^2e^{2\bar\phi}B_\ell'B_1').
\end{equation}
Variation of parameters gives the primitive \eqref{eq:growingprimitive}, with meromorphic power integration inherited from~\eqref{eq:intG0}. In the following formulas the variable is $X=r_{\mathrm a}^p/\mu-1$, and primes denote $X$ derivatives.

One term of $J_n^+$ resonates at $p=1$. With $s=n+\ell+1$, which is even, it is the term whose primitive becomes constant at $p=1$; it integrates with denominator $s(1/p-1)$, and its coefficient also vanishes at $p=1$. To find their ratio, differentiate the Legendre equation with respect to its degree, using the normalisation in~\eqref{eq:Fseries}. For nonnegative integer $j$, this shows that
\[
 \left.\partial_\nu F_\nu\right|_{\nu=j}-\tfrac12F_j\log(X^2-1)
\]
is a polynomial of degree at most $j-2$ and parity $j$, vanishing for $j=0,1$. Since $\left.\partial_pF_{j/p}(X)\right|_{p=1}=-j\left.\partial_\nu F_\nu(X)\right|_{\nu=j}$, only these logarithms can contribute to the coefficient of $X^{-1}$ in the differentiated integrand at $p=1$. Writing $[X^{-1}]$ for the coefficient of $X^{-1}$ in the large-$X$ expansion, for an odd polynomial $g$ we have $[X^{-1}]\,g(X)\log(X^2-1)=-2\int_0^1g(t)\,\dd t$, and $(X^2-1)F_n'F_\ell'$ is odd, so
\[
 [X^{-1}]\left.\partial_p\bigl[(X^2-1)F_{n/p}'F_{\ell/p}'F_{1/p}'\bigr]\right|_{p=1}
 =s\int_0^1(t^2-1)F_n'F_\ell'\,\dd t .
\]
Dividing by the derivative $-s$ of the denominator gives the added constant $-\int_0^1(t^2-1)F_n'F_\ell'\,\dd t$, which is the lower limit $1$ in~\eqref{eq:endconstant}. In closed form, the antiderivative
\begin{equation}
 \frac{(X^2-1)[-n(n+1)F_nF_\ell'+\ell(\ell+1)F_n'F_\ell]}
       {\ell(\ell+1)-n(n+1)}
\end{equation}
vanishes at $X=1$, so it equals $J_n^+(1,X)$, and its value at $0$ is the added constant. The decaying primitive has no resonance because $\ell-n$ is odd. At $p=1$, the two primitives give $h_{\ell-1}=-2(\ell+1)\mu^{\ell+1}F_\ell$ and $h_{\ell+1}=2\ell\mu^{\ell+1}F_\ell$, so the radial component of~\eqref{eq:restoring_dictionary} vanishes there.

Normalising the leading scalar term to $r^\ell$ gives \eqref{eq:physicalcolumn}. Without the finite constant, the result would not agree with the inner solution selected at generic dimension by the recurrence~\eqref{eq:recurrence}.

\subsection{Reconstruction of the higher-multipole response}\label{sec:app_higher}
We reconstruct the scalar response from the master variable and gauge residual of Section~\ref{sec:decoupled}, at $p=1$ with $\lambda=\ell(\ell+1)$, the notation~\eqref{eq:xy} and primes denoting $x$ derivatives. Solving~\eqref{eq:reduced} for $\Psi$ and $w_\pm$, and hence $w_r,w_\Omega$ by~\eqref{eq:wpm}, we first choose a representative (subscript $\mathrm{rep}$) with $v=e=0$, for which
\begin{align}
 \psi_{\rm rep}&=\Psi-w_\Omega/(2x),\qquad f_{\rm rep}=-2w_\Omega,\notag\\
 (\lambda-2)c_{\rm rep}&=-4(\psi_{\rm rep}'+2x\psi_{\rm rep}/\Delta)
 -\Delta w_r'-2xw_\Omega'+(4-\lambda-4/\Delta)w_\Omega .
\end{align}
Its gauge residual is corrected to the pair $w_r,w_\Omega$ above by the pure-gauge column~\eqref{eq:coordinate}, with $\xi_r=\ell Z_--(\ell+1)Z_+$ and $\xi_\Omega=(Z_-+Z_+)/x$, where, with $\mathcal N=-4\psi_{\rm rep}+2xc_{\rm rep}+(4x-2/x)w_\Omega$,
\begin{equation}
 L_{\ell-1}Z_-=\frac{\mathcal N}{2\ell+1},\qquad L_{\ell+1}Z_+=-\frac{\mathcal N}{2\ell+1}.
\end{equation}
The two Legendre operators differ by a constant, $L_{\ell-1}-L_{\ell+1}=2(2\ell+1)$, so their product $\mathcal P=L_{\ell-1}L_{\ell+1}=L_\ell^2-2L_\ell-4\lambda$ has inverse $(\Rop_{\ell+1}-\Rop_{\ell-1})/[2(2\ell+1)]$. (Operator products act to their right, inverse products denote composition of Green's functions, $\Delta^{-1}$ multiplies by $1/\Delta$, and braces denote the anticommutator.) Combining the restoring functions gives $\xi_r=\mathcal P^{-1}L_\ell\mathcal N$. The scalar response is $\psi_{\rm rep}+\xi_r/\Delta$, by the first entry of~\eqref{eq:coordinate}; expressing $\Psi$ and $w_\pm$ through the Green's functions~\eqref{eq:legendreGF}, with the normalisation~\eqref{eq:GK}, gives
\begin{equation}\label{eq:reconstruction_operator}
 K_\ell=\Mop_\ell-4(\mathcal P^{-1}\Delta^{-1}+\Delta^{-1}\mathcal P^{-1})
 +8\Delta^{-1}\mathcal P^{-1}\{3x^2-1,L_\ell-1\}\mathcal P^{-1}\Delta^{-1}.
\end{equation}
The second inverse product reduces by the same constant difference of Legendre operators:
\begin{equation}\label{eq:composition_reduction}
 \mathcal P^{-1}\{3x^2-1,L_\ell-1\}\mathcal P^{-1}
 =\frac{\mathcal C_{\ell+1}-\mathcal C_{\ell-1}}{2(2\ell+1)},
\end{equation}
as follows by multiplying both sides by $\mathcal P$ and using $L_{\ell-1}-L_{\ell+1}=2(2\ell+1)$ and $L_{\ell-1}+L_{\ell+1}=2(L_\ell-1)$: the difference of the two products with middle factor $3x^2-1$ is the anticommutator. Substituting this identity and $\mathcal P^{-1}$ into~\eqref{eq:reconstruction_operator} gives~\eqref{eq:Khigher}. Because these are identities between inverses, not only between differential operators, they require vanishing boundary forms. At the origin they vanish because the polynomial inner solutions of degrees $\ell\pm1$ have the same parity, so both their values or both their derivatives vanish there; at the upper endpoint the boundary forms decay.

\section{Dimensional inputs and force completion for the dipole}\label{sec:app_dipole}
\subsection{Exact dimensional coefficients}
We give the two generic-dimensional coefficients used in Section~\ref{sec:exceptional} to fix the dipole constants, and the finite algorithm that produces them. At generic $p$, with $\lambda=p+1$ and the row and column of $e$ deleted, we take three independent null vectors $\boldsymbol v_i$ of $\mathsf H_0(1,p)$ and one of $\mathsf H_0(3,p)$, with $\mathsf H_0$ as in~\eqref{eq:powerH}, so that the flat exponents are $j_i=1,1,1,3$, and build inner columns from $\beta_i=j_i$ and outer columns from $\beta_i=-j_i-p$ with~\eqref{eq:recurrence}. Their cross-pairing, the matrix $\mathcal W$ of~\eqref{eq:Green}, is $\mathcal W_{ij}=(-j_i-j_j-p)\boldsymbol v_i^T\mathsf K_0\boldsymbol v_j$. Setting the radius to one, so that $z_R=\mu$, and writing $[\mu^k]$ for the coefficient of $\mu^k$, the coefficient $T_{1,k}$ of the normalised dipole response~\eqref{eq:Tdef}, $T_1=\sum_kT_{1,k}z_R^k$, is
\begin{equation}
 T_{1,k}(p)=-\frac{2(p+1)(p+2)}p\,[\mu^k]\bigl[U_{\rm in}(1)\,\mathcal W^{-T}U_{\rm out}(1)^T\bigr]_{\psi\psi},
\end{equation}
where the suffix $\psi\psi$ extracts the scalar entry.
We define the following numerator and denominator polynomials in $p$:
\begin{align}
 \mathsf p_4={}&45p^8+1437p^7-27474p^6+22316p^5+243384p^4\notag\\
 &+189568p^3-129792p^2-138240p-18432,\notag\\
 \mathsf d_4={}&8(p-6)(p-4)(p-2)^2(p+4)(p+6)(3p-2)(3p+2)(3p+4),\\
 \mathsf p_6={}&216675p^{13}-2284470p^{12}+56135241p^{11}-64575702p^{10}\notag\\
 &-676183432p^9-358290448p^8+1384329104p^7+1411648160p^6\notag\\
 &-524465920p^5-1070609920p^4-234983424p^3+140894208p^2\notag\\
 &+57212928p+3538944,\notag\\
 \mathsf d_6={}&48(p-6)(p-4)(p-2)^3(p+4)(p+6)\notag\\
 &\times(3p-4)(3p-2)(3p+2)(3p+4)(5p-2)(5p+2)(5p+4).
\end{align}
The exact rational results, including the second-order coefficient $T_{1,2}$ as a check, are
\begin{equation}
 T_{1,2}(p)=-\frac{3(p+1)(p^2+14p+8)}{2(p-2)(p+4)(p+6)},
 \qquad
 T_{1,4}(p)=\frac{(p+1)\mathsf p_4(p)}{\mathsf d_4(p)},\quad
 T_{1,6}(p)=\frac{(p+1)\mathsf p_6(p)}{\mathsf d_6(p)}.
\end{equation}
Both denominators are nonzero at $p=1$. Substitution gives \eqref{eq:dipcoeff} and $T_{1,2}(1)=69/35$, the value used in~\eqref{eq:leading_sum}.

\subsection{Light-source dipole Green's function}\label{sec:app_lightdipole}
We construct the light-source part $K_L$ of the dipole Green's function, whose diagonal~\eqref{eq:lightdip} is used in Section~\ref{sec:dipole}. For the remainder of this appendix, $p=1$, $\lambda=2$, $x=r/\mu$, $y=R/\mu$ and $\Delta=x^2-1$, with primes denoting $x$ derivatives. At $\ell=1$, inverting~\eqref{eq:wpm} gives $w_r=w_--2w_+$ and $w_\Omega=x(w_-+w_+)$, and the Ward identities~\eqref{eq:reduced} read $L_0w_-=-4T/(3\Delta)$ and $L_2w_+=4T/(3\Delta)$.

For an arbitrary source $T$, start from the representative~\eqref{eq:diprep}, with $\psi_{\rm rep}=c_{\rm rep}=0$. The pure-gauge column~\eqref{eq:coordinate} that brings its residual to the required value has $\xi_r=Z_--2Z_+$ and $\xi_\Omega=(Z_-+Z_+)/x$, where
\begin{equation}\label{eq:dipZ}
 3L_0Z_-=\mathsf A_0w_-+\mathsf A_2w_++(3x^2-1)T,\qquad
 3L_2Z_+=\mathsf B_0w_-+\mathsf B_2w_++T,
\end{equation}
with the first-order operators
\begin{equation}
\begin{aligned}
 \mathsf A_0&=3x^2-1-x\Delta(9x^2-5)\partial_x,&\qquad \mathsf A_2&=-4(3x^2-2)-4x\Delta\partial_x,\\
 \mathsf B_0&=1+x\Delta\partial_x,&\qquad \mathsf B_2&=1-3x^2+x\Delta\partial_x .
\end{aligned}
\end{equation}
The full fields are then
\[
 \psi=\frac{\xi_r}{\Delta},\qquad f=f_{\rm rep}+2\xi_r',\qquad
 v=v_{\rm rep}+\frac{\xi_r+\Delta\xi_\Omega'}{x},\qquad c=\frac{2x\xi_r}{\Delta}-2\xi_\Omega .
\]
Eliminating $w_\pm''$ with the Ward identities and $Z_\pm''$ with~\eqref{eq:dipZ}, one finds that $E_f=E_v=E_c=0$ (Appendix~\ref{sec:app_rows}), that $E_\psi=T$, and that the residual of these fields is $w_r,w_\Omega$, for any source $T$, with $w_\pm$ satisfying the Ward identities. For $T=w_\pm=0$, the representative vanishes and the reconstruction reduces to the pure-gauge columns~\eqref{eq:coordinate} discussed in Section~\ref{sec:dipole}.

For $T=\delta(x-y)$ and the light-source residual, i.e.\ the terms in $\Rop_0$ and $\Rop_2$ of~\eqref{eq:dipward}, we invert~\eqref{eq:dipZ} with the Legendre Green's functions $\Rop_0$ and $\Rop_2$ of~\eqref{eq:legendreGF}. This produces compositions that generalise $\mathcal C_n$. For a first-order operator $\mathsf O=\mathsf O_0(t)+\mathsf O_1(t)\partial_t$, acting on the second factor, and $x<y$,
\begin{align}\label{eq:gencomposition}
 [\Rop_a\mathsf O\Rop_b](x,y)={}&Q_a(x)Q_b(y)\int_0^xP_a\,\mathsf OP_b\,\dd t
 +P_a(x)Q_b(y)\int_x^yQ_a\,\mathsf OP_b\,\dd t\notag\\
 &+P_a(x)P_b(y)\int_y^\infty Q_a\,\mathsf OQ_b\,\dd t,
\end{align}
where unlabelled functions are evaluated at $t$, and the case $x>y$ follows by splitting at $y$ and $x$. The integrals use the zero-constant power integration of~\eqref{eq:composition}, with the upper limit contributing its continued value zero. For the operators appearing here, the large-$t$ integrands contain no $t^{-1}$ term, so this integration generates no resonant logarithm. For $\mathsf O=3t^2-1$ and $a=b=n$, \eqref{eq:gencomposition} is $\mathcal C_n$ of~\eqref{eq:composition}. The light-source Green's function is
\begin{align}
 K_L(x,y)={}&\frac{-4[\Rop_0\mathsf A_0\Rop_0]+4[\Rop_0\mathsf A_2\Rop_2]+8[\Rop_2\mathsf B_0\Rop_0]-8[\Rop_2\mathsf B_2\Rop_2]}{9\Delta_x\Delta_y}\notag\\
 &+\frac{(3y^2-1)\Rop_0(x,y)-2\Rop_2(x,y)}{3\Delta_x},\label{eq:KL}
\end{align}
with every composition evaluated at $(x,y)$. On the diagonal, with $P_0=1$, $P_2=(3x^2-1)/2$ and the corresponding $Q_n$, all degree-zero compositions and derivative terms combine into the elementary terms of~\eqref{eq:lightdip}, leaving only $I_2$ of~\eqref{eq:In}.

Reciprocity can be shown before evaluating the integrals. With the formal adjoint taken in $\dd x$, $(\mathsf O_0+\mathsf O_1\partial_x)^\dagger=\mathsf O_0-\partial_x(\mathsf O_1\,\cdot\,)$, direct differentiation gives
\[
 \mathsf A_0-\mathsf A_0^\dagger=[L_0,k_0],\qquad
 \mathsf B_2-\mathsf B_2^\dagger=[L_2,k_2],\qquad
 \mathsf A_2=4\mathsf B_0^\dagger,\qquad
 \tfrac12\mathsf A_2=L_0k-kL_2,
\]
with $k_0=-9x^4/4+5x^2/2$, $k_2=x^2/2$ and $k=-\Delta/2$. Inserting these into the antisymmetric part of~\eqref{eq:KL}, the factors $L_n$ cancel adjacent Green's functions, and the remaining coefficients of $\Rop_0$ and $\Rop_2$ are the differences between their values at $x$ and $y$ of $4(k-k_0)-3\Delta(3x^2-1)=-1$ and $4(k-2k_2)+6\Delta=-4$, respectively. Both are constant, so both differences vanish. The boundary terms of these integrations by parts vanish as well: the degree-zero and degree-two inner solutions are even, with zero derivative at the origin, and the continued boundary terms at infinity have no constant part. Hence $K_L(x,y)=K_L(y,x)$.

\subsection{Heavy-force reconstruction}\label{sec:app_heavyforce}
We now construct the heavy-force term $-8\xi_H/\Delta^2$ of~\eqref{eq:dipfinal}, with the conventions of Appendix~\ref{sec:app_lightdipole}. Factoring out the common source strength $-4/(3\Delta_y)$ from the heavy-force part of~\eqref{eq:dipward}, its residual is $w_-=Q_0$, $w_+=0$, i.e.\ $w_r=Q_0$ and $w_\Omega=xQ_0$, and the representative~\eqref{eq:diprep} has
\[
 f_{\rm rep}=-\frac{3x^2-1}{2},\qquad
 v_{\rm rep}=-\frac{3\Delta}{4}.
\]
We add the pure-gauge column~\eqref{eq:coordinate} with $\xi_r=\xi_H$ and $\xi_\Omega=\xi_{\Omega,H}$, where
\begin{equation}\label{eq:heavyforce_displacement}
 \xi_H=\frac{(7-9x^2)Q_0+9x^3+18x}{36},\qquad
 \xi_{\Omega,H}=\frac{18x^3+(27x^2-13)Q_0+9x}{72x}.
\end{equation} (Equivalently, $\xi_H=Z_--2Z_+$ and $\xi_{\Omega,H}=(Z_-+Z_+)/x$, with $Z_\pm$ particular solutions of~\eqref{eq:dipZ} for $w_-=Q_0$, $w_+=0$ and $T=0$.) For the sum of the representative and this pure-gauge column, substitution into the definitions of $w_r,w_\Omega$ gives $Q_0$ and $xQ_0$, and substitution into the field equations of Appendix~\ref{sec:app_rows} gives $E_\psi=E_f=E_v=E_c=0$ away from the source. The scalar entry $\xi_H/\Delta$, with the source factor restored, gives the term $-4\xi_H(x)/(3\Delta_x\Delta_y)$ used in Section~\ref{sec:dipole}.

\section{Angular-tail bounds and dimensional continuation}\label{sec:app_tail}
This appendix proves Lemma~\ref{lem:tail} and establishes the continuation domain used in Section~\ref{sec:angular} to identify the subtracted sum~\eqref{eq:sumrule} with the EFT.

\subsection{Uniform estimate near the physical dimension}\label{sec:app_uniform}
We fix a weak-field order $k$, and let $T_{\ell,k}$ and $A_{{\rm loc},k}$ be the weak-field coefficients in~\eqref{eq:tail}, with $A_r,A_t$ as in Section~\ref{sec:local}. We prove Lemma~\ref{lem:tail}, controlling both the radial integrals and the angular factors uniformly in complex $p$ near one.

\paragraph{The leading local term.} We use the logarithmic radius $t=\log(r/R)$ and define $\widehat\psi$ by $\psi=e^{-pt/2}\widehat\psi$. After angular reduction and the rescaling above, the scalar part of the quadratic action associated with~\eqref{eq:scalar_source} involves, up to a constant factor, the operator
\begin{equation}
 -\partial_t A_r\partial_t+\nu^2A_t+\frac{p^2}{4}(A_r-A_t)+\frac p2\partial_tA_r,\qquad \nu=\ell+p/2 .
\end{equation}
There is no term linear in $\nu$. Freezing $A_r,A_t$ at the source gives the leading coincident inverse $1/(2\nu\sqrt{A_rA_t})$, hence $A_{\rm loc}$ after the normalisation~\eqref{eq:Tdef}. We now bound the corrections from the varying coefficients and from the coupling to the metric.
\begin{enumerate}[leftmargin=*]
\item \emph{Radial integrals.} At fixed $k$, the Born expansion of Section~\ref{sec:prescription} has finitely many insertion chains. Each vertex coefficient is a finite sum of analytic functions of $p$ times $e^{-apt}$, with integer $0\leq a\leq k$. For $\ell$ large compared with $k$, the flat Green's functions~\eqref{eq:G0j} of degrees $\ell,\ell\pm2$ make every radial integral absolutely convergent at both ends.
\item \emph{The angular factors are under control.} For real $p$ near one, sphere integration gives a positive Gram matrix on the Cartesian tensor sector, with respect to which the angular matrix $\mathsf K_0^{-1}\mathsf H_0(0)$ of Appendix~\ref{sec:app_recurrence} is self-adjoint. Its normalised entries and spectral projectors extend analytically to a complex neighbourhood of $p=1$, and the explicit matrix formulas give uniform bounds there, for the projectors and the inverse kinetic form alike. The angular matrix satisfies a cubic polynomial with eigenvalues $(\ell+\sigma)(\ell+\sigma+p)$, $\sigma=-2,0,2$. Subtracting $\ell(\ell+p)\Id$ leaves an $O(\nu)$ matrix. Each projector has a quadratic numerator, and its denominator is a product of two of the eigenvalue gaps $4\nu-4$, $4\nu+4$ and $8\nu$, so it is bounded below by a positive multiple of $|\nu|^2$ for large $\ell$, uniformly near $p=1$. The projectors therefore add no powers of $\nu$.
\item \emph{Mixing with the metric is suppressed.} For a chain of $s$ vertices, let $d_i\in\{0,1,2\}$ count derivatives at vertex $i=1,\ldots,s$. Rescaling integration coordinates by $|\nu|$ gives $|\nu|^{-1}$ per propagator and integration, and at most $|\nu|$ per derivative: the unnormalised size is $O(|\nu|^{\sum_i d_i-2s-1})$. A scalar-to-scalar chain passing through the metric has at least two mixing vertices, each with at most one derivative. Its normalised contribution is therefore $O(\nu^{-2})$.
\item \emph{A term of order $1/\nu$ would spoil the estimate, but it cancels.} Only scalar principal-part chains can contribute at order one or $1/\nu$. Their constant coefficients sum to $A_{\rm loc}$. The first Taylor correction contains one factor of an integration coordinate. Reflecting all integration coordinates preserves the free scalar Green's functions and the even derivative count, but reverses that factor. Its integral therefore vanishes identically in complex $p$.
\item \emph{The remaining terms are bounded uniformly.} Let $t_i$ be the logarithmic radius at vertex $i$, set $\tau_i=|\nu|t_i$, and fix the external coordinates $\tau_0=\tau_{s+1}=0$. For large $\ell$, there are constants $B_k,c_k>0$ and a nonnegative integer $b_k$ such that the rescaled Taylor remainder is bounded by
\begin{equation}
 B_k(1+\textstyle\sum|\tau_i|)^{b_k}
       e^{-c_k\sum|\tau_{i+1}-\tau_i|},
\end{equation}
which is integrable and gives a uniform $O(\nu^{-2})$ remainder. Insertions of the derivative-free terms have the same suppression. In the weak action form, only internal propagator edges can be twice differentiated. Their delta functions collapse integrations, preserving the counting and reflection argument without producing a $\delta(0)$.
\end{enumerate}

Together with the pole cancellation in Sections~\ref{sec:continuation} and~\ref{sec:exceptional}, this proves Lemma~\ref{lem:tail}.

\subsection{Continuation from the convergence strip}\label{sec:app_continuation}
The comparison with the original EFT integral starts in the strip $-1<\Re p<0$. There, the light-point behaviour $|\mathbf x-\mathbf x_2|^{-p}$ or softer is continuous at coincidence, and the mode sum~\eqref{eq:modesum}, a Gegenbauer expansion, converges absolutely. This expansion uses the same angular measure and normalisations as dimensional Fourier integration. The radial power integrals~\eqref{eq:intG0} and the contact term~\eqref{eq:contact} were already matched before coincidence. Auxiliary analytic indices or local Taylor subtractions may be retained at the heavy point and infinity during this comparison. On an open set in the strip, the coincident field integral therefore equals its convergent modal sum. Subtracting $A_{\rm loc}$ leaves this equality unchanged, since $\sum_\ell\omega_\ell=0$ in the strip by~\eqref{eq:weightsum}.

Continuing this equality to the physical dimension requires a connected domain linking this strip to $p=1$, with a summable bound throughout. At fixed $k$, we choose a bounded path in $-1<\Re p<2$ from a regular point of $(-1,0)$ to $1$, avoiding $p=0$ and the poles of the finitely many background and vertex coefficients. The possible positive background poles, from the denominators in~\eqref{eq:Arn}, are $p=2/(n-1)$ for positive even $n\leq k$, none equal to one.

On a compact neighbourhood of this path, vertex coefficients and exponential rates are bounded, while the decay rates of the free radial Green's functions grow uniformly with $\ell$. The estimates of Appendix~\ref{sec:app_uniform} therefore still apply after raising the angular cutoff. The finitely many modes below it have meromorphic fixed-order coefficients. Further detours avoid their nonremovable poles. Their apparent poles at $p=1$ are removable by the results of Sections~\ref{sec:continuation} and~\ref{sec:exceptional}.

We choose a connected open neighbourhood of the resulting path whose compact closure avoids these poles and lies in $\Re p<2$. Compactness gives a uniform gap from $\Re p=2$, so, since $\omega_\ell=O(\ell^{\Re p-1})$ by~\eqref{eq:omegal}, the weighted subtracted terms, $O(\ell^{\Re p-3})$, have a summable majorant independent of $p$ there. Their sum is therefore analytic in the neighbourhood, and it agrees with the original integral where the neighbourhood meets the strip. Uniqueness of continuation carries that equality to $p=1$. The neighbourhood and bounds may depend on $k$, so this establishes the coefficientwise identification stated in~\eqref{eq:sumrule}.

\section{Ancillary files and checks}\label{sec:app_checks}
Several steps of the derivation can be tested independently, and the code in the ancillary files does so. At the level of the operator, we regenerate the Hessian through sixth order in $\mu$, and we compare the Born expansion built from flat propagators with the recurrence~\eqref{eq:recurrence} through second order; this tests whether the recurrence indeed implements the point-source prescription~\eqref{eq:intG0}. We also check the algebraic and differential identities used for the radial solutions, keeping their parameters symbolic. For the dipole, the code checks the reconstruction of Appendix~\ref{sec:app_lightdipole} for arbitrary $T$ (\texttt{check\_low\_local.py}), the formal adjoint identities and algebraic reciprocity cancellations (\texttt{check\_low\_adjoint.py}), and the dipole Green's function against the generic-dimensional recurrence at distinct radii (\texttt{low\_kernel\_series.py}).

The dimensional limit provides a more stringent test. For $\ell=0,1,2,3,4$ and through sixth order, we check that the negative Laurent powers cancel in every perturbation component present in the sector, and we compare the finite scalar Green's functions at distinct radial arguments, logarithms included, with the closed forms of Sections~\ref{sec:higher} and~\ref{sec:exceptional}. Finally, we check the angular-matrix and projector identities used in Appendix~\ref{sec:app_tail}, as well as the potential coefficients~\eqref{eq:PNpredictions}, including the conversion to physical masses. The guide in the ancillary files explains how to reproduce all of these calculations.

The two exact dipole coefficients~\eqref{eq:dipcoeff} have a different role: they are inputs to the argument rather than tests of it, and they are computed from the finite recurrence of Appendix~\ref{sec:app_dipole} at generic dimension.

\phantomsection

\end{document}